%% file: ff4lbbh.tex
\documentclass[onecolumn,preprintnumbers,aps,superscriptaddress,nofootinbib,prd]{revtex4-2}
\pdfoutput=1
\usepackage{amsmath,amssymb,nccmath,graphicx,color,hyperref}
\usepackage{scalefnt}
\definecolor{darkblue}{rgb}{0.0,0.0,0.7}
\definecolor{nicered}{rgb}{0.7,0.1,0.1}
\definecolor{nicegreen}{rgb}{0.0,0.4,0.0}
\hypersetup{colorlinks,citecolor=nicegreen,linkcolor=nicered,urlcolor=nicered}
\graphicspath{ {fig/} }
\makeatletter
\newcommand{\oset}[3][0ex]{%
  \mathrel{\mathop{#3}\limits^{
    \vbox to#1{\kern-2\ex@
    \hbox{$\scriptstyle#2$}\vss}}}}
\makeatother

\newcommand{\ep}{\epsilon}
\newcommand{\coupl}[1]{\textcolor{darkblue}{#1}}
\newcommand{\CS}{\textcolor{darkblue}}
\newcommand{\pole}[1]{\textcolor{darkblue}{#1}}
\newcommand{\logpow}[1]{\textcolor{darkblue}{#1}}
\newcommand{\mint}[1]{\scalebox{0.85}{$#1$\,}}

\newcommand{\FDiag}[2]{
\begin{minipage}{0.12\textwidth}
$\includegraphics[width=\textwidth]{#2}\hspace*{-15ex}\raisebox{-1.5ex}{\CS{#1}}\hfill$
\end{minipage}
\hspace*{2ex}
} 

\allowdisplaybreaks[1]

\begin{document}

\preprint{MSUHEP-22-014, P3H-22-034, TTP22-020, SI-HEP-2022-06}

\title{\texorpdfstring
{The $Hb\bar{b}$ vertex at four loops and hard matching\\
coefficients in SCET for various currents}
{The Hbb vertex at four loops and hard matching coefficients in SCET for various currents}}

\author{Amlan Chakraborty}
\affiliation{Department of Physics and Astronomy, Michigan State University,
East Lansing, Michigan 48824, USA}
\affiliation{The Institute of Mathematical Sciences, HBNI, Taramani, Chennai 600113, India}

\author{Tobias Huber}
\affiliation{Naturwissenschaftlich-Technische Fakult\"{a}t, Universit\"{a}t Siegen,
Walter-Flex-Str.\ 3, 57068 Siegen, Germany}

\author{Roman N. Lee}
\affiliation{Budker Institute of Nuclear Physics, 630090 Novosibirsk, Russia}

\author{Andreas von Manteuffel}
\affiliation{Department of Physics and Astronomy, Michigan State University,
East Lansing, Michigan 48824, USA}

\author{Robert~M.~Schabinger}
\affiliation{Department of Physics and Astronomy, Michigan State University,
East Lansing, Michigan 48824, USA}

\author{Alexander V. Smirnov}
\affiliation{Research Computing Center, Moscow State University,
119991 Moscow, Russia}
\affiliation{Moscow Center for Fundamental and Applied Mathematics,
119992 Moscow, Russia}

\author{Vladimir A. Smirnov}
\affiliation{Skobeltsyn Institute of Nuclear Physics of Moscow State University,
119991 Moscow, Russia}
\affiliation{Moscow Center for Fundamental and Applied Mathematics,
119992 Moscow, Russia}

\author{Matthias Steinhauser}
\affiliation{Institut f{\"u}r Theoretische Teilchenphysik,
Karlsruhe Institute of Technology (KIT),
76128 Karlsruhe, Germany}

\begin{abstract}
  \noindent
  We compute the four-loop corrections to the Higgs-bottom vertex within massless
  QCD and present analytic results for all color structures.
  The infrared poles of the renormalized form factor agree with the predicted
  four-loop pattern.
  Furthermore, we use the results for the Higgs-bottom, photon-quark,
  and Higgs-gluon form factors to provide hard matching coefficients in
  soft-collinear effective theory up to four-loop accuracy.
\end{abstract}

\maketitle

\section{Introduction}

In the next decade the Higgs boson will play a central role in many of the
analyses performed with data taken at the general purpose experiments ATLAS and
CMS at the Large Hadron Collider (LHC) at CERN.  Improved analysis tools
will increase the precision of observables, motivating new calculations on the
theory side in order to match the uncertainties of the experiments.

The dominant channel for Higgs boson production is via gluon fusion.  Although
in the Standard Model (SM) the contribution from bottom quark annihilation is
only at the percent level it might be important in extended models with
an enhanced coupling of the Higgs boson to bottom quarks. In the SM,
state-of-the-art for the inclusive production rate $b\bar{b} \to H +X$, with
$X$ being any hadronic state, is next-to-next-to-next-to-leading order (N$^3$LO)~\cite{Duhr:2019kwi,Duhr:2020kzd} (see
refs.~\cite{Dicus:1998hs,Balazs:1998sb,Harlander:2003ai} for the NNLO
corrections).
An important ingredient in the analysis of ref.~\cite{Duhr:2020kzd}
is the three-loop Higgs-bottom form factor which has been computed
in ref.~\cite{Gehrmann:2014vha} and cross checked in refs.~\cite{Lee:2017mip,Duhr:2019kwi}.
In this work we provide results for the four-loop corrections to the
Higgs-bottom form factor, which constitutes a building block for the Higgs
boson production cross section and the differential rate
of Higgs boson decays to bottom quarks at N$^4$LO.

There are two different ways to view Higgs boson production in bottom-antibottom
quark annihilation. In the so-called five-flavour scheme, the bottom quark is
considered as a massless parton which is part of the proton.
In the four-flavour scheme, the first step is the production of two (massive) bottom and
antibottom quark pairs via gluon splitting. Afterwards a bottom and an antibottom quark annihilate
to produce the Higgs boson. From the technical point of view the four-flavour
scheme is more involved and, in fact, in this approach only NLO corrections
are available~\cite{Dittmaier:2003ej,Dawson:2003kb,Wiesemann:2014ioa}, which
are of the same order in the strong coupling as the N$^3$LO corrections
available in the five-flavour scheme~\cite{Duhr:2019kwi,Duhr:2020kzd}.

An attractive framework to describe high-energy cross sections is based on soft-collinear effective theory (SCET)~\cite{Bauer:2000ew,Bauer:2000yr,Bauer:2001ct,Bauer:2001yt,Beneke:2002ph,Beneke:2002ni,Becher:2014oda}, which can be used to separate the various scales present in a process and to perform a resummation of potentially large logarithms. The matching of SCET to QCD is achieved with the help of so-called hard matching coefficients which can be extracted from the massless
form factors computed in QCD.
In this work, we calculate matching coefficients from results for the four-loop $Hb\bar{b}$, $\gamma^\ast q\bar{q}$ and $Hgg$ form factors.
An important aspect of this derivation is an explicit check of the
prediction~\cite{Mueller:1979ih,Collins:1980ih,Sen:1981sd,Magnea:1990zb,Sterman:2002qn,Ravindran:2004mb,Bern:2005iz,Moch:2005id,Moch:2005tm,Ravindran:2006cg,Dixon:2008gr,Becher:2009cu,Becher:2009qa} for the infrared poles of the renormalized $Hb\bar{b}$ form factor
using recent results for the four-loop cusp~\cite{Lee:2019zop,Bruser:2019auj,Henn:2019swt,vonManteuffel:2020vjv} and quark collinear~\cite{vonManteuffel:2020vjv,Agarwal:2021zft} anomalous dimensions.

The outline of the paper is as follows.
In section~\ref{sec::fb}, we discuss the bare Higgs-bottom form factor and present our result for the four-loop term.
For the Higgs-bottom, photon-quark and Higgs-gluon form factors, we discuss the
ultraviolet (UV) renormalization in section~\ref{sec::uv} and the infrared (IR) subtractions in section~\ref{sec::ir}.
The SCET hard matching coefficients for all three form factors are presented
in section~\ref{sec::match}.
We conclude in section~\ref{sec::concl}.

\section{\texorpdfstring
{The $H b\bar{b}$ form factor at four loops} 
{The Hbb form factor at four loops}}
\label{sec::fb}

\begin{figure} 
  \begin{center}
  \FDiag{$C_F^4$}{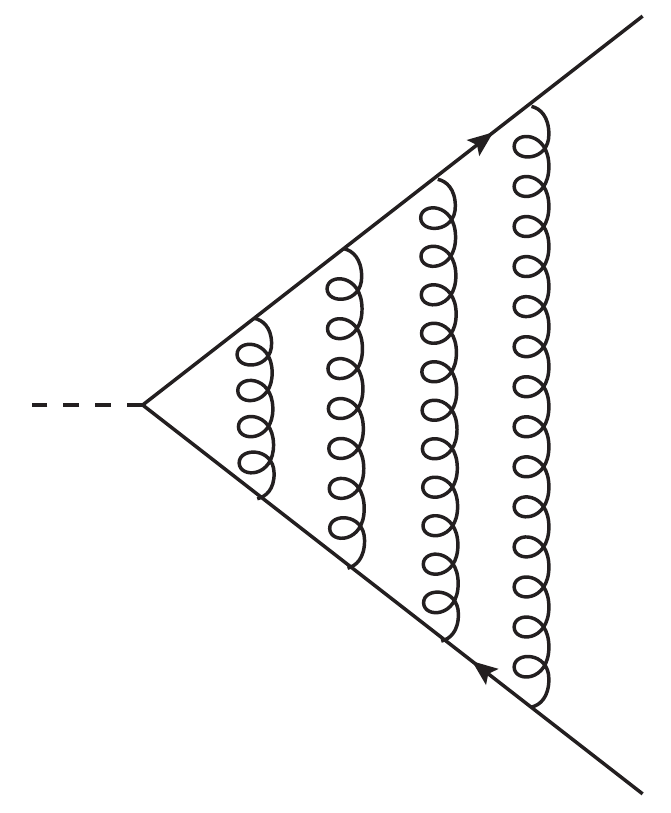}
  \FDiag{$C_F^3 C_A$}{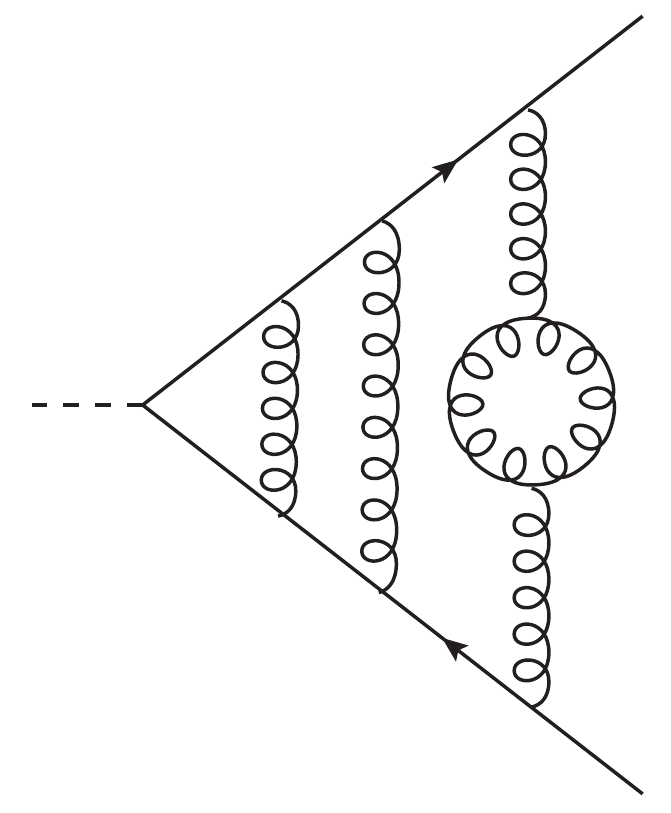}
  \FDiag{$C_F^2 C_A^2$}{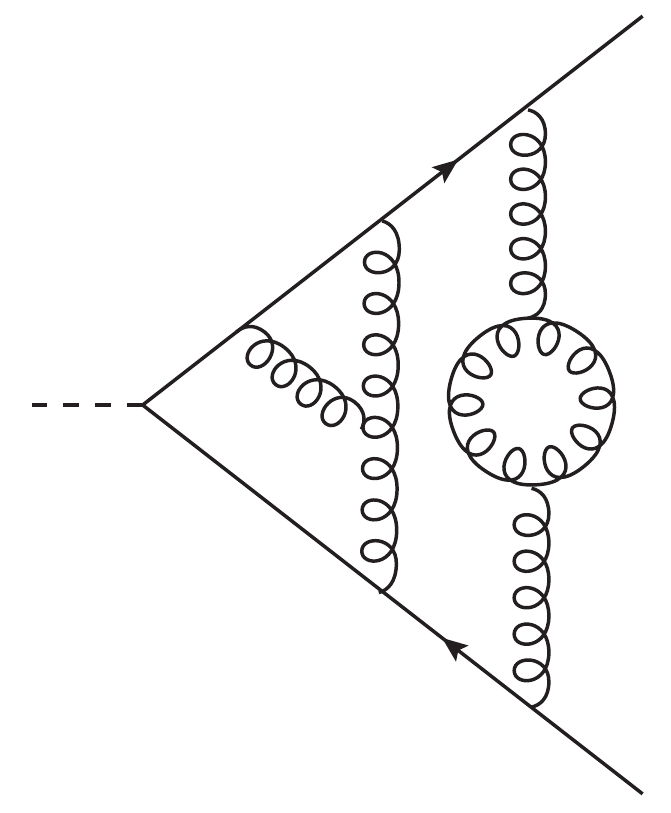}
  \FDiag{$C_F C_A^3$}{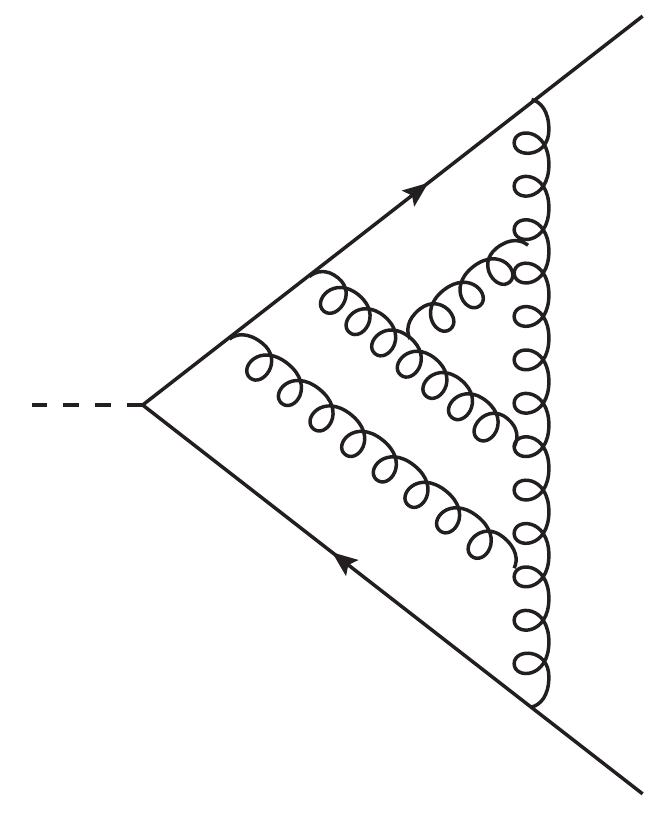}
  \FDiag{$d^{abcd}_F d^{abcd}_A/N_F$}{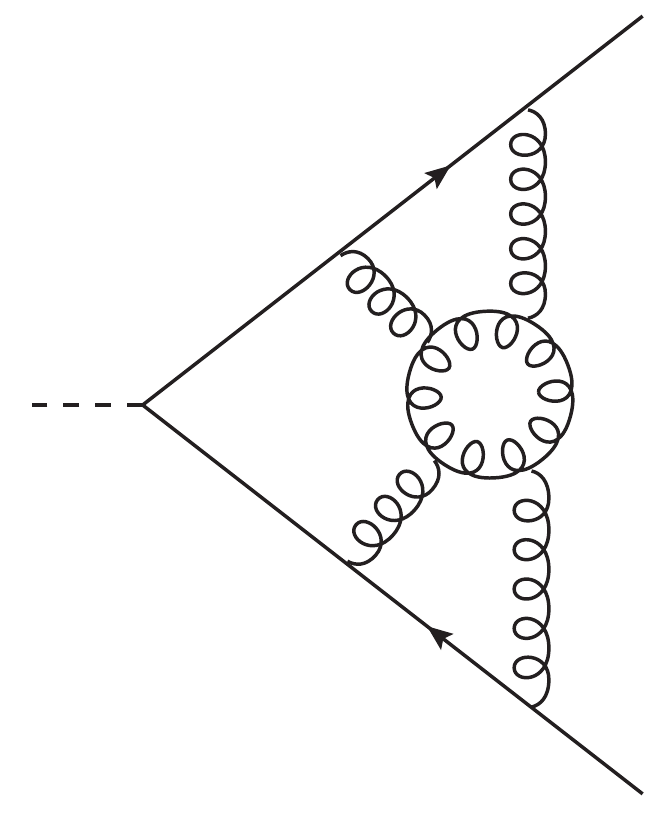}
  \FDiag{$n_f C_F^3$}{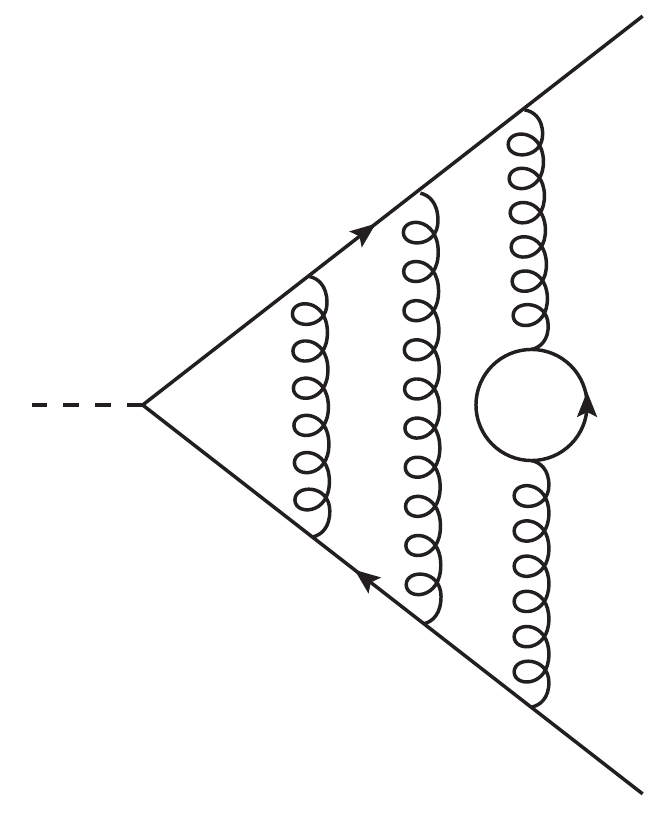}
  \\[1ex]
  \FDiag{$n_f C_F^2 C_A$}{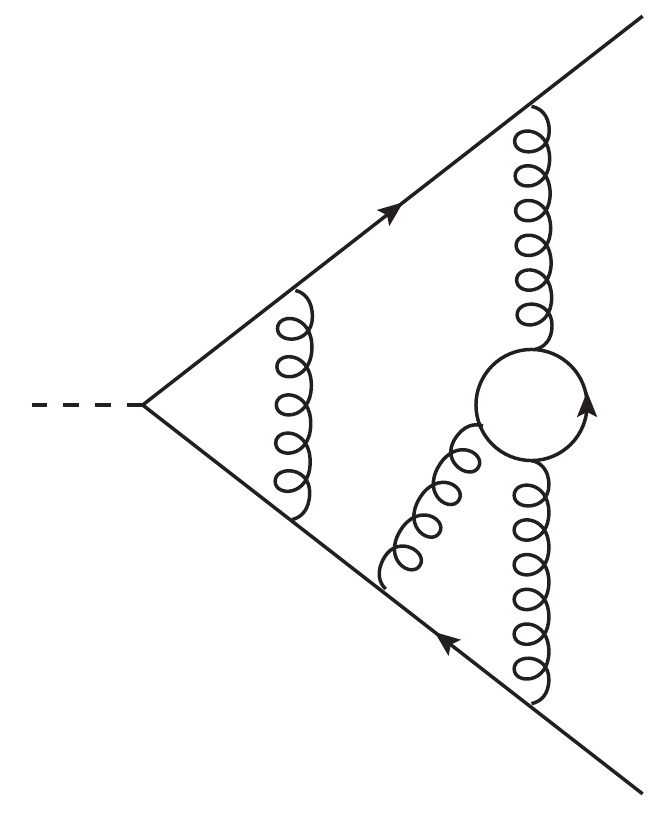}
  \FDiag{$n_f C_F C_A^2$}{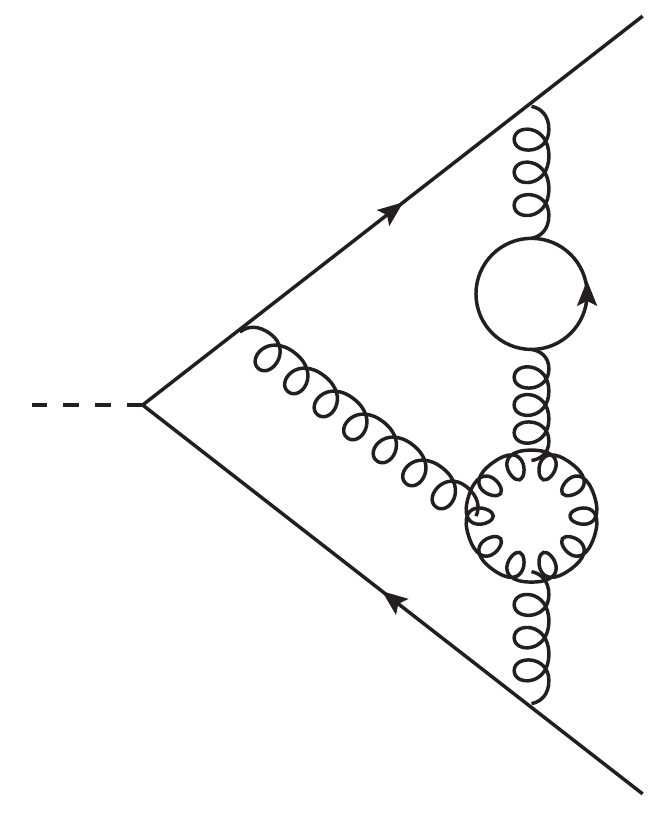}
  \FDiag{$n_f (d^{abcd}_F)^2/N_F$}{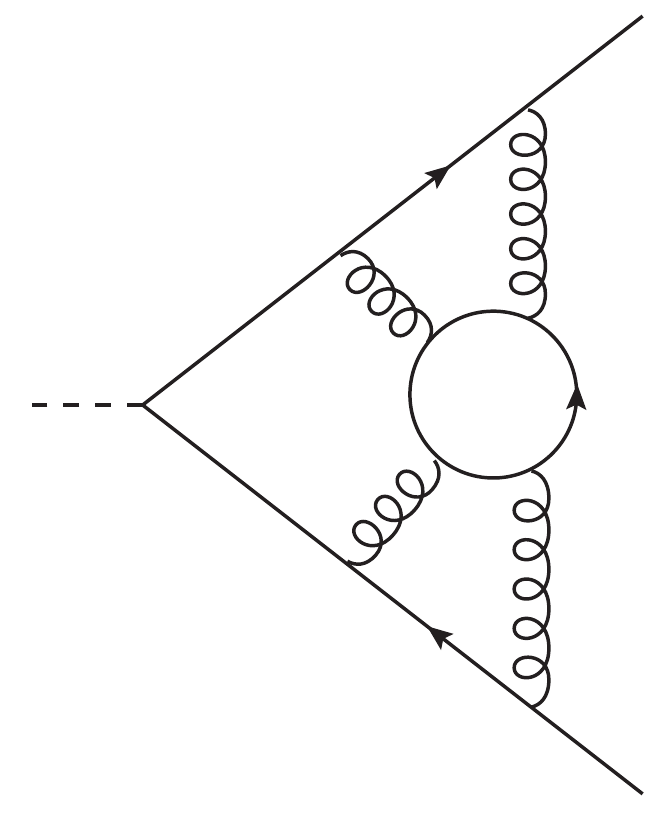}
  \FDiag{$n_f^2 C_F^2$}{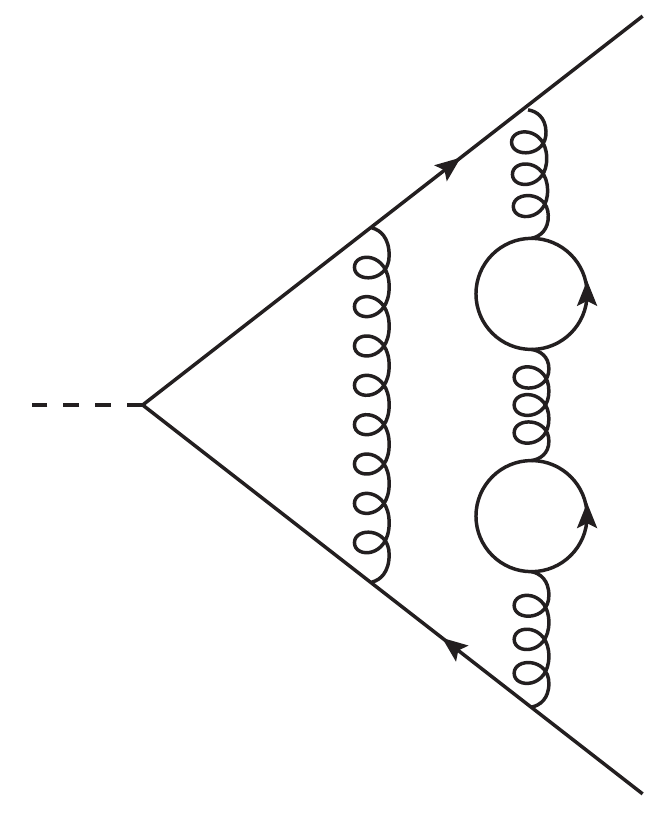}
  \FDiag{$n_f^2 C_F C_A$}{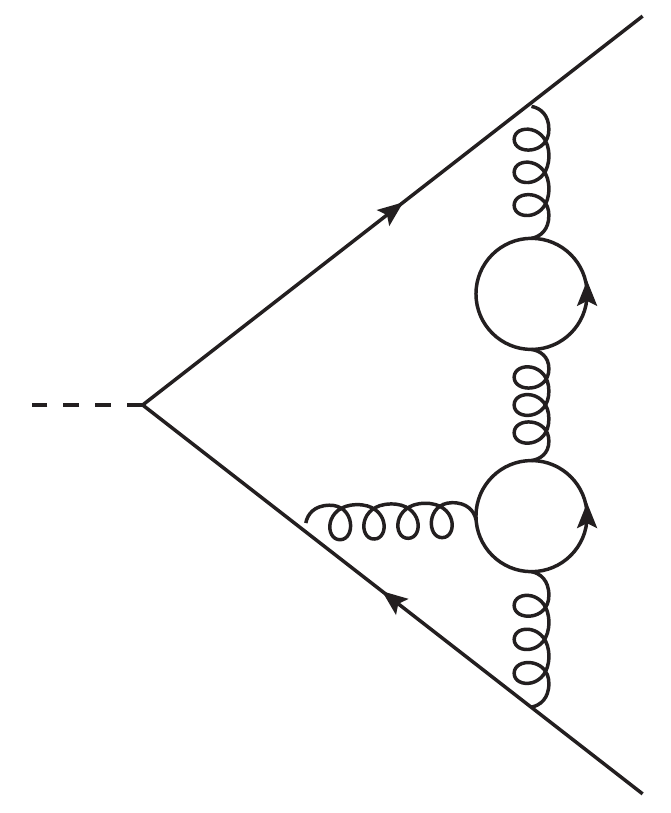}
  \FDiag{$n_f^3 C_F$}{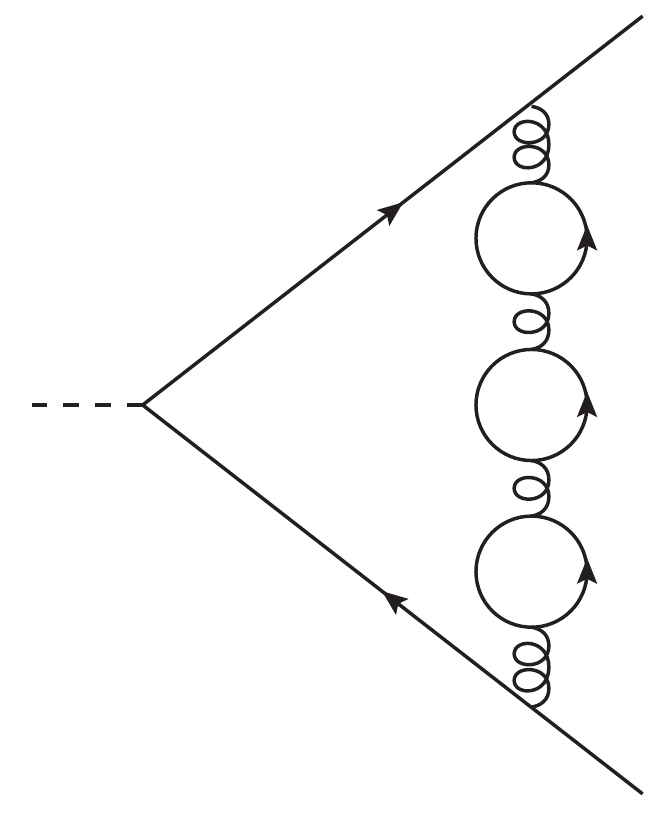}
  \caption{\label{fig::diags}Sample Feynman diagrams contributing to the
      $Hb\bar{b}$ form factor $F_b$ at the four-loop order.
      Solid and curly lines represent massless quarks and gluons, respectively.
      Both planar and non-planar diagrams contribute.}
  \end{center}
\end{figure}

We define the Higgs-bottom form factor via
\begin{align}
\label{eq::FFb}
  \mathcal{F}_b(q^2) &= -\frac{1}{2q^2}
  \mbox{Tr}\left( q_2\!\!\!\!\!/\,\,\, \Gamma_b \, q_1\!\!\!\!\!/\,\,\,
  \right)
  \,,
\end{align}
where $\Gamma_b$ is the Higgs-bottom vertex function, $q_1$ and $q_2$ are the
incoming quark and antiquark
momenta and $q=q_1+q_2$ is the momentum of the Higgs boson.
Sample Feynman diagrams contributing to $F_b$ at the four-loop level are shown in
fig.~\ref{fig::diags}. 
We employ conventional dimensional regularization to regularize UV
and IR divergences, and for the number of space-time dimensions $d$
we use $d=4-2\ep$.
Two- and three-loop corrections to $F_b$ have been computed in
refs.~\cite{Anastasiou:2011qx}
and~\cite{Gehrmann:2014vha}, respectively. The extension of the three-loop
result
up to order $\ep^2$ and the four-loop results with two and three closed
fermion loops have been computed in~\cite{Lee:2017mip}. In this work
we complete the four-loop corrections by computing the contributions of
all remaining color factors.

The bare form factor is conveniently parametrized in terms of the bare strong
coupling constant and the bare Yukawa coupling $y_0=m_{b,0}/v$ where
$m_{b,0}$ and $v$ are the bare bottom quark mass and Higgs vacuum expectation value.
The perturbative expansion of this bare form factor reads
\begin{align}
  \mathcal{F}_b &= y_0\left[1 +
  \sum_{n\ge1} 
  a_0^n
  \left(\frac{\mu_0^2}{-q^2-i0} \right)^{n\ep}
  S_\ep^{n} F_b^{(n)}
  \right]
  \,,
  \label{eq::FFbexp}
\end{align}
where
\begin{align}
    S_\ep=e^{-\ep \gamma}(4\pi)^\ep,\qquad
    a_0 = \frac{\alpha_s^0}{4\pi},
\end{align}
$\alpha_s^0$ is the bare strong coupling, $\mu_0$ the 't Hooft scale,
$\gamma\approx 0.577216$ Euler's constant, and the $-i0$ description
fixes the branch cut ambiguity for $q^2 > 0$.
We stress that we keep a non-zero bottom mass only in the Yukawa coupling
and treat the bottom quark as a massless particle otherwise. 
$F_b^{(n)}$ develops poles up to $1/\ep^{2n}$. However, non-trivial 
information specific to some loop order is contained only in the $1/\ep^2$ poles and higher order $\ep$ terms. All higher poles, $1/\ep^{2n}, \dots, 1/\ep^3$,
are fixed from the lower-loop contributions.  In fact, the $1/\ep^2$ poles are determined by the (universal)
cusp anomalous dimension and the $1/\ep$ poles are determined by the collinear anomalous dimension. 

We generate the Feynman diagrams with {\tt Qgraf}~\cite{Nogueira:1991ex} and employ
{\tt Form\;4}~\cite{Kuipers:2012rf} to express the form factor in terms of unreduced
scalar Feynman integrals.
For the color algebra we use {\tt Color.h}~\cite{vanRitbergen:1998pn}, which conveniently
produces a result that is valid for a general simple Lie algebra.
We find the color structures listed in fig.~\ref{fig::diags},
where $C_R$ is the quadratic Casimir operator and
$d_R^{abcd}$ is the fully symmetrical tensor originating from the
trace over four generators,
with $R=F,A$ for the fundamental and adjoint representation, respectively.
Further, $N_F$ the dimension of the fundamental representation
and $n_f$ is the number of light quarks.
For a $SU(N_c)$ gauge group the relevant invariants or color factors
read
\begin{align}
\label{eq:casimir}
C_F &= (N_c^2-1)/(2 N_c),\nonumber\\
C_A &= N_c,\nonumber\\
d_F^{abcd}d_F^{abcd} / N_F &= (18-6N_c^2+N_c^4)(N_c^2-1)/(96N_c^3),\nonumber\\
d_F^{abcd}d_A^{abcd} / N_F &= (N_c^2-1)(N_c^2+6)/48.
\end{align}
We note that diagrams where the Higgs couples to a closed quark loop
do not contribute.
This is clear from the fact that the Yukawa coupling requires a helicity flip
and all quarks are massless in our calculation.
For the same reason, there are also no $d^{abc}_Fd^{abc}_F$ contributions,
which are present in the case of the $\gamma^\ast q\bar{q}$ form factor~\cite{Lee:2021uqq}.

The computation of massless four-loop form factor integrals requires advanced
techniques both for the reduction to master integrals but also for the
computation of the latter.  For the reduction it is essential to have at hand
an efficient program.
For our calculation we use
\texttt{Reduze\;2}~\cite{vonManteuffel:2012np} together with the code
\texttt{Finred} employing finite field arithmetic and further techniques from
\cite{vonManteuffel:2014ixa,Gluza:2010ws,Larsen:2015ped,Boehm:2017wjc,Lee:2014tja,Bitoun:2017nre,Agarwal:2020dye}.

For the master integrals two complementary methods
are applied. The first one uses finite master
integrals~\cite{vonManteuffel:2014qoa,vonManteuffel:2015gxa,Schabinger:2018dyi}
in $d_0-2\epsilon$ dimensions where $d_0=4,6,\ldots$ and applies the program
{\tt HyperInt}~\cite{Panzer:2014caa} in cases where the corresponding Feynman
parametric representation can be rendered linearly
reducible~\cite{Brown:2008um,Brown:2009ta}.
In this approach, the actual integration is applied to individual master integrals.
The second method considers all master integrals of a given integral family at
the same time. In a first step one of the massless external legs is made
massive. Choosing $q_1^2\not=0$ it is possible to define $x=q_1^2/q^2$, where
$q^2$ is the virtuality of the Higgs boson. Our aim is the computation of the
master integrals for $x=0$. On the other hand, for $x=1$ the vertex integrals
turn into massless two-point functions, which are well studied in the
literature~\cite{Baikov:2010hf,Lee:2011jt}.  It is indeed possible to use the powerful method of
differential equations~\cite{Kotikov:1990kg,Bern:1993kr,Remiddi:1997ny,Gehrmann:1999as,Henn:2013pwa, Lee:2014ioa} to transport the
information from $x=1$ to $x=0$. For more details on this approach we refer to
ref.~\cite{Lee:2019zop}.

We obtain for the bare four-loop form factor
\begin{align}
\input{eq/fb.tex}
\,.
\label{eq::FFb4res}
\end{align}
Our result is expressed in terms of regular zeta values $\zeta_2$, $\zeta_3$,
$\zeta_5$, $\zeta_7$ and one multiple zeta value
\begin{equation}
    \zeta_{5,3} = \sum_{m=1}^\infty \sum_{n=1}^{m-1} \frac{1}{m^5 n^3} \approx 0.0377076729848\,.
\end{equation}
As expected for a generic four-loop form factor in QCD, the leading pole is
$1/\ep^8$ and the finite part has transcendental weight up to 8.

We have performed the following checks on our result in eq.~\eqref{eq::FFb4res}
for the bare four-loop form factor $F_b^{(4)}$.
First, we have recalculated the known $n_f^2$ and $n_f^3$ contributions
and found agreement with the results of ref.~\cite{Lee:2017mip};
all other terms are new.
For the leading color contribution we have performed two independent
calculations and verified that the results agree.
Furthermore, we have checked that all poles $1/\ep^8, \ldots, 1/\ep$ 
agree with predictions derived from known anomalous dimensions through to
four-loop order and lower loop contributions through to transcendental weight 8.
This is a strong check of our result
and a confirmation of the literature expression for the quark collinear
anomalous dimension.
Details for this derivation will be given in the sections below.
For the sake of completeness, we have also recalculated the lower loop
bare form factors $F_b^{(1)}$, $F_b^{(2)}$, $F_b^{(3)}$ through to weight 8 and
found full agreement with the results of ref.~\cite{Lee:2017mip}.
We also extracted the maximal transcendental weight 8 part and find that
it coincides with that of the $\gamma^\ast q\bar{q}$ form factor~\cite{Lee:2021uqq,Lee:2022nhh}
(and, after modifying the Casimir operators such that quarks and
gluons are in the adjoint color representation, also with that of the $Hgg$
form factor~\cite{Lee:2021uqq,Lee:2022nhh}).

\section{\texorpdfstring
{UV renormalization for $Hb\bar{b}$, $\gamma^\ast q\bar{q}$ and $Hgg$ form factors}
{UV renormalization for Hbb, Aqq and Hgg form factors}}
\label{sec::uv}

In this section, we discuss the UV renormalization of different form factors,
focusing first on the $Hb\bar{b}$ form factor.
We perform UV renormalization in the $\overline{\rm MS}$ scheme and replace the bare couplings $a_0$ and $y_0$ by the renormalized couplings $a$ and $y$ with
\begin{align}
\label{eq::adef}
S_\ep \mu_0^{2\ep} a_0 &= Z_a \mu^{2\ep} a,\\
\label{eq::ydef}
y_0 &= Z_m y\,.
\end{align}
With the $\beta$ function and the quark anomalous dimension $\gamma^m$
\begin{alignat}{2}
\label{eq::betaexp}
    \beta(a) &= - a \frac{\mathrm{d}\ln Z_a}{\mathrm{d} \ln \mu^2} &&= -a^2 \sum_{n=0}^\infty a^n \beta_n\,, \\
\label{eq::gammamexp}
    \gamma^m(a) &= - \frac{\mathrm{d}\ln Z_m}{\mathrm{d} \ln \mu^2} &&= -a \sum_{n=0}^\infty a^n \gamma^m_n\,,
\end{alignat}
one has from \eqref{eq::adef} for the renormalized coupling $\mathrm{d}a/\mathrm{d}\ln\mu^2 = \beta - a\ep$ and thus
\begin{align}
\frac{\mathrm{d}\ln Z_a}{\mathrm{d}a} &= -\frac{\beta}{a(\beta-a\ep)},\\
\frac{\mathrm{d}\ln Z_m}{\mathrm{d}a} &= -\frac{\gamma^m}{\beta - a \ep}\,.
\end{align}
Solving these differential equations perturbatively results in
\begin{align}
   Z_a &= 1 + \coupl{a} \left( - \frac{\beta_0}{\pole{\ep}}  \right)
   + \coupl{a^2} \left(\frac{\beta_0^2}{\pole{\ep^2}}  - \frac{\beta_1}{2\pole{\ep}}   \right)
   + \coupl{a^3} \left(\frac{7\beta_0\beta_1}{6\pole{\ep^2}} - \frac{\beta_0^3}{\pole{\ep^3}}
     -\frac{\beta_2}{3\pole{\ep}}  \right)
   + \coupl{a^4} \left( \frac{\beta_0^4}{\pole{\ep^4}} - \frac{23\beta_0^2\beta_1}{12\pole{\ep^3}}
     + \frac{9\beta_1^2 + 20\beta_0\beta_2}{24\pole{\ep^2}} - \frac{\beta_3}{4\pole{\ep}} 
     \right) \notag\\ &
   + \mathcal{O}(\coupl{a^5})\,,
   \\
   Z_m &= 1 + \coupl{a} \left( -\frac{\gamma^m_0}{\pole{\ep}} \right)
   + \coupl{a^2} \left( \frac{\gamma^m_0 (\beta_0 + \gamma^m_0)}{2\pole{\ep^2}}
     - \frac{\gamma^m_1}{2\pole{\ep}} \right)
   + \coupl{a^3} \left( -\frac{\gamma^m_0(\beta_0+\gamma^m_0)(2\beta_0+\gamma^m_0)}{6\pole{\ep^3}}
     + \frac{2\beta_1\gamma^m_0+2\beta_0\gamma^m_1+3\gamma^m_0\gamma^m_1}{6\pole{\ep^2}}
\right. \notag\\ & \left.
     + \frac{\gamma^m_2}{3\pole{\ep}} \right)
   + \coupl{a^4} \left( \frac{\gamma^m_0(\beta_0 + \gamma^m_0)(2\beta_0 + \gamma^m_0)
       (3\beta_0 + \gamma^m_0)}{24\pole{\ep^4}}
     + \frac{-6\beta_0\beta_1\gamma^m_0 - 4\beta_1 (\gamma^m_0)^2 - 3\beta_0^2\gamma^m_1
      - 7\beta_0\gamma^m_0\gamma^m_1 - 3(\gamma^m_0)^2\gamma^m_1}{12\pole{\ep^3}}
\right. \notag\\ & \left.
     + \frac{6\beta_2\gamma^m_0 + 6\beta_1\gamma^m_1 + 3(\gamma^m_1)^2 + 6\beta_0\gamma^m_2
       + 8\gamma^m_0\gamma^m_2}{24\pole{\ep^2}}
     - \frac{\gamma^m_3}{4\pole{\ep}} \right)
   + \mathcal{O}(\coupl{a^5})\,.
\end{align}
The coefficients of the $\beta$ function read~\cite{vanRitbergen:1997va,Czakon:2004bu}
\begin{align}
\input{eq/beta.tex}
\intertext{and the coefficients of the quark mass anomalous dimension are given by~\cite{Chetyrkin:1997dh,Vermaseren:1997fq}}
\input{eq/mass.tex}
.
\end{align}

In addition to the $Hb\bar{b}$ form factor~\eqref{eq::FFb}, we consider also the
UV renormalization of the bare form factors for
the $\gamma^\ast q\bar{q}$ and $Hgg$ vertices
\begin{align}
\label{eq::FFq}
  \mathcal{F}_q &= -\frac{1}{4(1-\epsilon)q^2}
  \mbox{Tr}\left( q_2\!\!\!\!\!/\,\,\, \Gamma^\mu_q q_1\!\!\!\!\!/\,\,\,
    \gamma_\mu\right)
  \,,
 \\
\label{eq::FFg}
  \mathcal{F}_g &=
  \frac{\left(q_1\cdot q_2\,\,
      g_{\mu\nu}-q_{1,\mu}\,q_{2,\nu}-q_{1,\nu}\,q_{2,\mu}\right)}
  {2(1-\epsilon)}
  \Gamma^{\mu\nu}_g
  \,.
\end{align}
Here, the projections are applied to the  $\gamma^\ast q\bar{q}$ and
$Hgg$ vertex functions $\Gamma^{\mu}_q$ and $\Gamma^{\mu\nu}_g$.
The $Hgg$ interaction is taken in the infinite top-quark-mass limit, where it can be
described by the bare effective Lagrangian
\begin{align}
  \mathcal{L}_{\text{eff}}&= -\frac{\lambda_0}{4} \, H F^{\mu\nu}_a \, F_{a,\mu\nu} \, .\label{eq::LeffHgg}
\end{align}
The bare form factor $\mathcal{F}_g$ depends on the bare coupling $\lambda_0$,
which is renormalized according to
\begin{align}
\lambda_0 = Z_\lambda \lambda \, ,
\end{align}
where the renormalization constant $Z_\lambda$ is given to all orders by coefficients of the QCD $\beta$-function~\cite{Spiridonov:1984br},
\begin{align}
  Z_\lambda &= \frac{1}{1-\beta/(a\ep)} \notag\\
  &=
  1 - \coupl{a}\frac{\beta_0}{\pole{\ep}}
  + \coupl{a^2} \left( \frac{\beta_0^2}{\pole{\ep^2}}  - \frac{\beta_1}{\pole{\ep}} \right) 
 + \coupl{a^3} \left(-\frac{\beta_0^3}{\pole{\ep^3}} + \frac{2 \beta_0 \beta_1}{\pole{\ep^2}} - \frac{\beta_2}{\pole{\ep}} \right)
 + \coupl{a^4} \left(\frac{\beta_0^4}{\pole{\ep^4}} - \frac{3 \beta_0^2 \beta_1}{\pole{\ep^3}} + \frac{\beta_1^2 + 2 \beta_0 \beta_2}{\pole{\ep^2}} - \frac{\beta_3}{\pole{\ep}}
    \right) + \mathcal{O}(\coupl{a^5})
    \, .
\end{align}

In summary, we arrive at the renormalized form factors for the $Hb\bar{b}$, $\gamma^\ast q\bar{q}$ and $Hgg$ vertices, respectively,
\begin{alignat}{2}
  \mathcal{F}_b^{\text{ren}} &= y \, Z_m\left[1 +\sum_{n\ge1} 
  \left(a \, Z_a\right)^n \, e^{-n \ep L} \, F_b^{(n)},
  \right] 
  &= y F^\text{ren}_b
  \label{eq::FFbrengeneral}
  \\[1.0em]
  \mathcal{F}_q^{\text{ren}} &= 1 +\sum_{n\ge1} 
  \left(a \, Z_a\right)^n \, e^{-n \ep L} \, F_q^{(n)}
  &= F^\text{ren}_q
  \label{eq::FFqrengeneral}
  \,,
  \\[1.0em]
  \mathcal{F}_g^{\text{ren}} &= \lambda \, Z_\lambda\left[1 +\sum_{n\ge1} 
  \left(a \, Z_a\right)^n \, e^{-n \ep L} \, F_g^{(n)}
  \right]
  &= \lambda F^\text{ren}_g
  \, ,
  \label{eq::FFgrengeneral}
\end{alignat}
where
\begin{equation}
    L \equiv \ln\left(\frac{-q^2-i0}{\mu^2} \right)
\end{equation}
contains the dependence on the renormalization scale.

\section{\texorpdfstring
{IR subtraction for $Hb\bar{b}$, $\gamma^\ast q\bar{q}$ and $Hgg$ form factors}
{IR subtraction for Hbb, Aqq, Hgg form factors}}
\label{sec::ir}

We begin by considering a general UV renormalized scattering amplitude $\mathcal{M}^\text{ren}$
with IR poles in $\ep$.
These divergences shall be absorbed by introducing a quantity $\mathbf{Z}$ such that
\begin{align}
    \mathcal{M}^\text{fin} = \mathbf{Z}^{-1} \mathcal{M}^\text{ren},
\end{align}
where $\mathcal{M}^\text{fin}$ is finite for $\epsilon \to 0$.
For strongly interacting external states, $\mathcal{M}^\text{ren}$ and
$\mathcal{M}^\text{fin}$ are vectors in color space and $\mathbf{Z}$ is a matrix.
Interestingly, the matrix $\mathbf{Z}$ exhibits universal, process-independent features,
see \cite{Agarwal:2021ais} for a recent review.
Defining the anomalous dimension matrix $\mathbf{\Gamma}$ via
\begin{align}
\mathbf{\Gamma}(\mu,a) = -\mathbf{Z}^{-1}\frac{\mathrm{d}\mathbf{Z}}{\mathrm{d}\ln \mu} 
\end{align}
the matrix $\mathbf{Z}$ can be expressed as \cite{Becher:2009qa} 
\begin{align}
\ln \mathbf{Z} =
- \frac{1}{2}\int_0^a\!\! \frac{\mathrm{d}a'}{\beta(a')-\ep a'}
\left(\mathbf{\Gamma}(\mu,a')
-  \frac{1}{2}\int_0^{a'}\!\! \frac{\mathrm{d}a''\,\,\Gamma'(a'')}{\beta(a'')-\ep a''} \right)\,. \label{eq::lnZ}
\end{align}
Expansion in $a$ according to eq.~\eqref{eq::betaexp},
\begin{align}
\mathbf{\Gamma}(\mu,a) &= \sum_{n=1}^\infty a^n \mathbf{\Gamma}_n(\mu)\,,\\
\Gamma'(a) &= \frac{\mathrm{d}\mathbf{\Gamma}(\mu,a)}{\mathrm{d}\ln(\mu)}
 = \sum_{n=1}^\infty a^n \Gamma'_n\,,
\end{align}
and integration gives \cite{Becher:2019avh} 
\begin{align}
\label{eq::lnZexp}
\ln \mathbf{Z} &= \coupl{a} \bigg(\frac{\Gamma'_1}{4\pole{\ep^2}} + \frac{\mathbf{\Gamma}_1}{2\pole{\ep}}\big) + 
 \coupl{a^2} \bigg(
 - \frac{3 \beta_0 \Gamma'_1}{16 \pole{\ep^3}}
 + \frac{\Gamma'_2-4 \beta_0 \mathbf{\Gamma}_1}{16\pole{\ep^2}}
 + \frac{\mathbf{\Gamma}_2}{4\pole{\ep}}
 \bigg)
   + 
 \coupl{a^3} \bigg(
  \frac{11 \beta_0^2 \Gamma'_1}{72 \pole{\ep^4}}
 + \frac{
    12 \beta_0^2 \mathbf{\Gamma}_1 - 8 \beta_1 \Gamma'_1 - 5 \beta_0 \Gamma'_2}{72 \pole{\ep^3}}
\notag\\ &
+ \frac{\Gamma'_3 - 6 \beta_0 \mathbf{\Gamma}_2 -6 \beta_1 \mathbf{\Gamma}_1 
  }{36 \pole{\ep^2}}
 + \frac{\mathbf{\Gamma}_3}{6\pole{\ep}}
 \bigg)
+ \coupl{a^4} \bigg(
 - \frac{25 \beta_0^3 \Gamma'_1}{192 \pole{\ep^5}} 
 + \frac{\beta_0 (13 \beta_0 \Gamma'_2 + 40 \beta_1 \Gamma'_1 
   - 24 \beta_0^2 \mathbf{\Gamma}_1 )}{192 \pole{\ep^4}}
\notag\\ &
 + \frac{ - 7 \beta_0 \Gamma'_3 -  9 \beta_1 \Gamma'_2
   + 24 \beta_0^2 \mathbf{\Gamma}_2 - 15 \beta_2 \Gamma'_1 
   + 48 \beta_0 \beta_1 \mathbf{\Gamma}_1 }{192 \pole{\ep^3}}
 + \frac{ \Gamma'_4 - 8 \beta_0 \mathbf{\Gamma}_3  - 8 \beta_1 \mathbf{\Gamma}_2
   -8 \beta_2 \mathbf{\Gamma}_1 }{64 \pole{\ep^2}}
 + \frac{\mathbf{\Gamma}_4}{8 \pole{\ep}}
 \bigg)
 + \mathcal{O}(\coupl{a^5})\,.
\end{align}
Through to three loops, the matrices $\mathbf{\Gamma}_n$ and $\Gamma'_n$ 
are known~\cite{Aybat:2006wq,Aybat:2006mz,Becher:2009cu,Dixon:2009gx,Gardi:2009zv,Gardi:2009qi,Becher:2009qa,Almelid:2015jia} in terms of cusp and collinear anomalous dimensions,
depending only on the type of external state.
In particular, they contain a sum over so-called dipole contributions, each of which
is generated from color correlations of two external states.
Starting at three loops, also quadrupole contributions involving three or four external partons at a time appear~\cite{Almelid:2015jia,Henn:2016jdu,Becher:2019avh,Caola:2021rqz,Caola:2021izf} through the  anomalous dimension matrix $\mathbf{Z}$. At four loops, structural information about the matrix $\mathbf{Z}$ is available~\cite{Becher:2019avh,Agarwal:2020nyc,Agarwal:2021him,Falcioni:2021buo}, but its complete expression is not known yet.

For four loop form factors with only two colored external states,
there is only one dipole contribution and the structure simplifies significantly.
In particular, the matrix $\mathbf{Z}$ becomes diagonal and, see~\cite{Becher:2009qa},
\begin{align}
  \mathbf{Z} &= Z_r\,,\\
  \mathbf{\Gamma}_n &= - \Gamma^r_n \ln\left(\frac{\mu^2}{-q^2-i0}\right) - \gamma^r_n\,,\\
  \Gamma'_n &= -2\Gamma^r_n\,, \label{eq::ZGammadipole}
\end{align}
where $r=q,g$ denotes the type of external particle
and
\begin{align}
\Gamma^r(a) &= \sum_{n=1}^\infty a^n \Gamma^r_n\,,\\
\gamma^r(a) &= \sum_{n=1}^\infty a^n \gamma^r_n
\end{align}
are the cusp and collinear anomalous dimensions, respectively.
The coefficients of the cusp anomalous dimension through to four-loop order are~\cite{Henn:2019swt,vonManteuffel:2020vjv}
\begin{align}
\label{eq::cusp}
\input{eq/cusp.tex}
,
\intertext{where $R=F$ for $r=q$ and $R=A$ for $r=g$.
The collinear anomalous dimensions are known to four-loop order as well~\cite{vonManteuffel:2020vjv,Agarwal:2021zft}, and
the coefficients read}
\input{eq/colq.tex}
\label{eq::colq}
\intertext{for the quark and}
\input{eq/colg.tex}
\label{eq::colg}
\end{align}
for the gluon.
We note that the exact conventions used above differ slightly from \cite{Becher:2009qa},
and our notation for the cusp and collinear anomalous dimensions coincide with that of
refs.~\cite{vonManteuffel:2020vjv,Agarwal:2021zft}.
In particular, the quantity $\gamma^q_i$ used here is $-2$ times the quantity denoted by $\gamma^q_{i-1}$ in ref.~\cite{Becher:2009qa}.

With the above, the finite remainders for our form factors are obtained as
\begin{align}
    F^\text{fin}_b &= Z_q^{-1} F^\text{ren}_b, \label{eq::finremb}\\
    F^\text{fin}_q &= Z_q^{-1} F^\text{ren}_q, \label{eq::finremq}\\
    F^\text{fin}_g &= Z_g^{-1} F^\text{ren}_g, \label{eq::finremg}
\end{align}
We stress that the factor $Z_q$ is the same for the $\gamma^\ast q\bar{q}$ and $Hb\bar{b}$
form factors.
The fact that our explicit four-loop result for $F^\text{fin}_b$ derived from
eq.~\eqref{eq::FFb4res} indeed is finite for $\ep\to 0$ is a non-trivial
check of the pole subtraction framework and the involved anomalous
dimensions.

Finally, we note that the scheme considered here is a \emph{minimal} subtraction
of just the poles in $\ep$ for $\ln F^\text{ren}_r$,
order-by-order in the coupling $a$.
While the poles of $\ln F^\text{ren}_r$ are equal to the poles of $\ln Z_r$ and thus are
universal, due to exponentiation, the poles of
$F^\text{ren}_r$ at a given loop order are process dependent.
In particular, their prediction based on \eqref{eq::lnZexp} involves also higher order $\ep$
contributions from $F^\text{ren}_r$ at lower loops.

\section{\texorpdfstring
{Hard matching coefficients in SCET from $Hb\bar{b}$, $\gamma^\ast q\bar{q}$, and $Hgg$ form factors}
{Hard matching coefficients in SCET  from Hbb, Aqq, and Hgg form factors}}
\label{sec::match}

In physical problems with widely separated scales the perturbative expansion can be spoiled since powers of the coupling are accompanied by powers of logarithms of large scale ratios. In such cases, the large logarithms can be resummed to all orders in perturbation theory by means of renormalization-group techniques formulated in the language of effective field theory. For the calculation of cross sections and kinematic distributions in collider physics the appropriate framework is provided by SCET~\cite{Bauer:2000ew,Bauer:2000yr,Bauer:2001ct,Bauer:2001yt,Beneke:2002ph,Beneke:2002ni,Becher:2014oda}, which is used for instance in Drell-Yan and Higgs production for rapidity~\cite{Ahmed:2014uya,Ahmed:2014era,Li:2016axz,Li:2016ctv,Ebert:2017uel,Dulat:2018bfe,Ahmed:2020amh,Chen:2021vtu}, transverse-momentum~\cite{Catani:2013tia,Becher:2014tsa,Chen:2018pzu,Gutierrez-Reyes:2019vbx,Gutierrez-Reyes:2019rug,Scimemi:2019cmh,Billis:2021ecs,Camarda:2021ict} and thrust distributions~\cite{Abbate:2010xh,Monni:2011gb,Gao:2019mlt}, or for the treatment of threshold effects in deep-inelastic scattering, see \textit{e.g.}~\cite{Ajjath:2020sjk,Abele:2022wuy}. For all of these applications, the hard matching coefficients in SCET are required, which can be extracted from the form factors discussed above.

In dimensionally regularized SCET, the IR divergences in $F^\text{ren}_r$, $r=b,q,g$, become the UV poles of the bare matching coefficients.
In particular, performing the matching on-shell, loop integrals in SCET are scaleless and vanish, \textit{i.e.}\ their UV and IR poles cancel each other.
Furthermore, the IR poles must reproduce those of $F^\text{ren}_r$, and hence we obtain the renormalized matching coefficients $C^r$ by subtracting the IR poles of $F^\text{ren}_r$ through a multiplicative renormalization factor, which is precisely the procedure applied in eqs.~\eqref{eq::finremb}\,--\,\eqref{eq::finremg}. We can therefore define the SCET hard matching coefficients $C^{r}$ for $r=b,q,g$ and their perturbative expansion according to
\begin{align}
C^{r} &= \lim_{\ep\to 0} F^{\text{fin}}_r
= 1 + \sum_{n=1}^\infty a^n \, C^{r}_n\, .
\end{align}
The matching coefficients depend on the renormalization scale $\mu$
through the renormalized coupling $a=a(\mu^2)$ and the logarithm $L$.
Results through to three loops are available for $r=q,g$ in the literature~\cite{Gehrmann:2010ue}, while to the best of our knowledge the full $H b\bar{b}$ matching coefficient is presented here for the first time (the $L$-independent part through to three loops can be found in~\cite{Ebert:2017uel}).

We start with the matching coefficient for the $Hb\bar{b}$ form factor,
\begin{align}
\input{eq/cb.tex}
.
\end{align}
In the case of the $\gamma^\ast q\bar{q}$ form factor, the lower-loop results can be found in~\cite{Gehrmann:2010ue} and we use the same normalization here. The four-loop result can be extracted from  $F_q^{(4)}$ in ref.~\cite{Lee:2022nhh} and reads
\begin{align}
\input{eq/cq.tex}
.
\end{align}
Finally, we consider the $Hgg$ case. The lower-loop results can be found in~\cite{Gehrmann:2010ue} as well, and we find from the $Hgg$ form factor $F_g^{(4)}$ in ref.~\cite{Lee:2022nhh}
\begin{align}
\input{eq/cg.tex}
\end{align}
In $C^q_4$ and $C^g_4$, $d_F^{abc}$ is the fully symmetrical tensor originating from the trace over three generators, $N_A$ is the dimension of the adjoint representation, and $n_{q\gamma} =\textstyle \sum_{q^\prime}Q_{q^\prime}/Q_q$ is the charge-weighted sum over the quark flavours normalized to the charge of the external quark, see refs.~\cite{Lee:2021uqq,Lee:2022nhh} for details.

Taking $\mathrm{d}/\mathrm{d}\ln\mu$ of eqs.~\eqref{eq::finremb}\,--\,\eqref{eq::finremg} and using the IR structure presented in eqs.~\eqref{eq::lnZ}\,--\,\eqref{eq::ZGammadipole} one can derive a renormalization group equation (RGE) for the hard matching coefficients, which is used to resum logarithms of disparate scales in the SCET framework. The RGE assumes the expected generic form (see \textit{e.g.}~\cite{Becher:2006nr,Ahrens:2009cxz,Bell:2010mg})
\begin{align}
 \frac{\mathrm{d} \, C^r}{\mathrm{d} \ln \mu} &= \bigg[\Gamma^r(a) \, L - \gamma^r(a) -2 \, \mathcal{G}^r\bigg] \, C^r \, . \label{eq::RGE}
\end{align}
For a given particle species and color representation the RGE consists of two universal terms related to the renormalization properties of the SCET current, of which the cusp anomalous dimension $\Gamma^r$ controls the leading Sudakov double logarithms, while the collinear anomalous dimension $\gamma^r$ is responsible for resumming single logarithms. The third term $\mathcal{G}^r$, which also governs the single-logarithmic evolution, is related to the anomalous dimension of the QCD current and therefore is non-universal. We get $\mathcal{G}^q=0$ since the vector current is conserved, while for the scalar current this piece reads $\mathcal{G}^b=- (\mathrm{d}\ln Z_m)/(\mathrm{d} \ln \mu^2)=\gamma^m$. For the gluonic case we find quite analogously $\mathcal{G}^g=- (\mathrm{d}\ln Z_\lambda)/(\mathrm{d} \ln \mu^2)=a \, \mathrm{d}(\beta/a)/\mathrm{d}a$. We checked explicitly that all our matching coefficients satisfy~\eqref{eq::RGE} through to four-loop order.

\section{Conclusions}  
\label{sec::concl}

In this paper we computed the four-loop corrections
to the $H b\bar{b}$ vertex in massless QCD.
Our main result is the analytic expression for the bare
form factor presented in eq.~(\ref{eq::FFb4res}).
After renormalization of the strong coupling constant
and the Higgs-bottom Yukawa coupling, the infrared poles
agree with the form predicted in the literature and
confirm previous results for the cusp and quark collinear
anomalous dimensions.
In addition to the new results for the Higgs-bottom form factor,
we considered the previously published four-loop results
for the bare photon-quark and Higg-gluon form factors.
For all three cases, we employed $Z$ factors to minimally subtract
the IR poles from the renormalized form factors,
extracted the finite SCET hard matching coefficients,
and presented the analytic four-loop results.
Our results are available in plain text format in the ancillary
files on arXiv.

\bigskip
{\bf Acknowledgments.}
AvM and RMS gratefully acknowledge Erik Panzer for related collaborations.
This research was supported by the Deutsche Forschungsgemeinschaft (DFG,
German Research Foundation) under grant 396021762 — TRR 257 ``Particle Physics
Phenomenology after the Higgs Discovery'' and by the National Science
Foundation (NSF) under grant 2013859 ``Multi-loop amplitudes and precise
predictions for the LHC''.  The work of AVS and VAS was supported by the Ministry of Education and Science of the Russian Federation as part of the program of the Moscow Center for Fundamental and Applied Mathematics under agreement no. 075-15-2019-1621. The work of RNL is supported by the Russian Science Foundation, agreement no. 20-12-00205.
We acknowledge the High Performance Computing Center at Michigan State University for computing resources.  The Feynman diagrams were drawn with the help of {\tt
  Axodraw}~\cite{Vermaseren:1994je} and {\tt JaxoDraw}~\cite{Binosi:2003yf}.

\bibliography{ff4lbbh}

\end{document}

%% file: eq/fb.tex
F_b^{(4)} &= 
\CS{C_F^4} \bigg[
 \pole{\frac{1}{\ep^8}} \Big(
   \mfrac{2}{3}
 \Big)+ \pole{\frac{1}{\ep^6}} \Big(
  - \mfrac{4}{3}\zeta_2
  + \mfrac{8}{3}
 \Big)+ \pole{\frac{1}{\ep^5}} \Big(
  - \mfrac{272}{9}\zeta_3
  + \mint{12}\zeta_2
  + \mfrac{16}{3}
 \Big)+ \pole{\frac{1}{\ep^4}} \Big(
  - \mfrac{296}{15}\zeta_2^2
  - \mint{60}\zeta_3
  + \mfrac{80}{3}\zeta_2
  + \mfrac{68}{3}
 \Big)
 \notag\\ &
 + \pole{\frac{1}{\ep^3}} \Big(
  - \mfrac{3008}{15}\zeta_5
  + \mfrac{640}{9}\zeta_3 \zeta_2
  - \mint{12}\zeta_2^2
  - \mfrac{2336}{9}\zeta_3
  + \mfrac{340}{3}\zeta_2
  + \mint{52}
 \Big)+ \pole{\frac{1}{\ep^2}} \Big(
   \mfrac{19360}{27}\zeta_3^2
  - \mfrac{6784}{315}\zeta_2^3
  - \mint{1100}\zeta_5
  - \mint{480}\zeta_3 \zeta_2
  + \mfrac{118}{15}\zeta_2^2
 \notag\\ &
  + \mfrac{668}{9}\zeta_3
  + \mint{506}\zeta_2
  - \mfrac{254}{3}
 \Big)+ \pole{\frac{1}{\ep}} \Big(
  - \mfrac{14162}{21}\zeta_7
  + \mfrac{5792}{5}\zeta_5 \zeta_2
  + \mfrac{6208}{9}\zeta_3 \zeta_2^2
  - \mint{1180}\zeta_3^2
  - \mfrac{17326}{21}\zeta_2^3
  - \mfrac{113542}{15}\zeta_5
  - \mfrac{21398}{9}\zeta_3 \zeta_2
 \notag\\ &
  + \mfrac{4867}{5}\zeta_2^2
  + \mfrac{69733}{9}\zeta_3
  + \mfrac{5159}{2}\zeta_2
  - \mfrac{12707}{6}
 \Big)+  \Big(
  - \mfrac{32384}{15}\zeta_{5,3}
  + \mfrac{739328}{45}\zeta_5 \zeta_3
  - \mfrac{66392}{27}\zeta_3^2 \zeta_2
  + \mfrac{7486576}{7875}\zeta_2^4
  - \mfrac{47217}{2}\zeta_7
 \notag\\ &
  - \mfrac{31928}{5}\zeta_5 \zeta_2
  - \mfrac{11092}{5}\zeta_3 \zeta_2^2
  - \mfrac{284228}{27}\zeta_3^2
  - \mfrac{250138}{45}\zeta_2^3
  - \mfrac{392059}{15}\zeta_5
  - \mfrac{121270}{9}\zeta_3 \zeta_2
  + \mfrac{28514}{3}\zeta_2^2
  + \mfrac{212006}{3}\zeta_3
  + \mfrac{53859}{4}\zeta_2
 \notag\\ &
  - \mfrac{71295}{4}
 \Big)\bigg]
+ \CS{C_F^3 C_A} \bigg[
 \pole{\frac{1}{\ep^7}} \Big(
  - \mfrac{11}{3}
 \Big)+ \pole{\frac{1}{\ep^6}} \Big(
   \mint{2}\zeta_2
  - \mfrac{67}{9}
 \Big)+ \pole{\frac{1}{\ep^5}} \Big(
   \mint{26}\zeta_3
  - \mfrac{638}{27}
 \Big)+ \pole{\frac{1}{\ep^4}} \Big(
   \mfrac{78}{5}\zeta_2^2
  + \mfrac{2018}{9}\zeta_3
  - \mfrac{209}{3}\zeta_2
  - \mfrac{5704}{81}
 \Big)
 \notag\\ &
 + \pole{\frac{1}{\ep^3}} \Big(
   \mint{182}\zeta_5
  - \mint{96}\zeta_3 \zeta_2
  + \mfrac{7096}{45}\zeta_2^2
  + \mfrac{27823}{27}\zeta_3
  - \mfrac{3505}{9}\zeta_2
  - \mfrac{53285}{243}
 \Big)+ \pole{\frac{1}{\ep^2}} \Big(
  - \mint{854}\zeta_3^2
  + \mfrac{13976}{315}\zeta_2^3
  + \mfrac{71882}{45}\zeta_5
  + \mfrac{2272}{9}\zeta_3 \zeta_2
 \notag\\ &
  + \mfrac{50633}{135}\zeta_2^2
  + \mfrac{279512}{81}\zeta_3
  - \mfrac{96061}{54}\zeta_2
  - \mfrac{846803}{1458}
 \Big)+ \pole{\frac{1}{\ep}} \Big(
   \mint{6172}\zeta_7
  - \mfrac{11548}{5}\zeta_5 \zeta_2
  - \mfrac{10508}{15}\zeta_3 \zeta_2^2
  - \mfrac{214747}{27}\zeta_3^2
  + \mfrac{173389}{945}\zeta_2^3
 \notag\\ &
  + \mfrac{1950199}{135}\zeta_5
  + \mfrac{162041}{27}\zeta_3 \zeta_2
  + \mfrac{2971}{162}\zeta_2^2
  + \mfrac{938374}{243}\zeta_3
  - \mfrac{2756239}{324}\zeta_2
  + \mfrac{2800297}{8748}
 \Big)+  \Big(
   \mfrac{20948}{15}\zeta_{5,3}
  - \mfrac{18364}{5}\zeta_5 \zeta_3
  + \mfrac{3296}{9}\zeta_3^2 \zeta_2
 \notag\\ &
  + \mfrac{5472536}{2625}\zeta_2^4
  - \mfrac{933223}{56}\zeta_7
  + \mfrac{69307}{9}\zeta_5 \zeta_2
  - \mfrac{262058}{27}\zeta_3 \zeta_2^2
  - \mfrac{803111}{81}\zeta_3^2
  + \mfrac{49163803}{5670}\zeta_2^3
  + \mfrac{141880529}{1620}\zeta_5
  + \mfrac{7016987}{162}\zeta_3 \zeta_2
 \notag\\ &
  - \mfrac{9911939}{972}\zeta_2^2
  - \mfrac{219806467}{2916}\zeta_3
  - \mfrac{84811717}{1944}\zeta_2
  + \mfrac{1218736471}{52488}
 \Big)\bigg]
+ \CS{C_F^2 C_A^2} \bigg[
 \pole{\frac{1}{\ep^6}} \Big(
   \mfrac{4961}{648}
 \Big)+ \pole{\frac{1}{\ep^5}} \Big(
  - \mfrac{451}{54}\zeta_2
  + \mfrac{32665}{972}
 \Big)+ \pole{\frac{1}{\ep^4}} \Big(
   \mfrac{397}{90}\zeta_2^2
 \notag\\ &
  - \mfrac{7975}{54}\zeta_3
  + \mfrac{371}{36}\zeta_2
  + \mfrac{209435}{1944}
 \Big)+ \pole{\frac{1}{\ep^3}} \Big(
   \mfrac{272}{3}\zeta_5
  + \mfrac{293}{9}\zeta_3 \zeta_2
  - \mfrac{32843}{270}\zeta_2^2
  - \mfrac{242806}{243}\zeta_3
  + \mfrac{117517}{486}\zeta_2
  + \mfrac{1236571}{4374}
 \Big)+ \pole{\frac{1}{\ep^2}} \Big(
   \mfrac{6065}{18}\zeta_3^2
 \notag\\ &
  + \mfrac{67988}{945}\zeta_2^3
  - \mfrac{21511}{18}\zeta_5
  + \mfrac{25544}{81}\zeta_3 \zeta_2
  - \mfrac{84901}{135}\zeta_2^2
  - \mfrac{3787327}{729}\zeta_3
  + \mfrac{5009207}{2916}\zeta_2
  + \mfrac{6218239}{13122}
 \Big)+ \pole{\frac{1}{\ep}} \Big(
  - \mfrac{38323}{18}\zeta_7
  + \mfrac{7921}{9}\zeta_5 \zeta_2
 \notag\\ &
  + \mfrac{35767}{135}\zeta_3 \zeta_2^2
  + \mfrac{590333}{81}\zeta_3^2
  - \mfrac{74927}{315}\zeta_2^3
  - \mfrac{6243851}{810}\zeta_5
  - \mfrac{83435}{27}\zeta_3 \zeta_2
  - \mfrac{516463}{243}\zeta_2^2
  - \mfrac{14345651}{729}\zeta_3
  + \mfrac{28502297}{2916}\zeta_2
  - \mfrac{15113323}{13122}
 \Big)
 \notag\\ &
 +  \Big(
   \mfrac{54254}{45}\zeta_{5,3}
  - \mfrac{45281}{9}\zeta_5 \zeta_3
  + \mfrac{38482}{27}\zeta_3^2 \zeta_2
  - \mfrac{21476059}{15750}\zeta_2^4
  - \mfrac{2480981}{144}\zeta_7
  + \mfrac{1053169}{270}\zeta_5 \zeta_2
  + \mfrac{3609022}{405}\zeta_3 \zeta_2^2
  + \mfrac{33473012}{729}\zeta_3^2
 \notag\\ &
  - \mfrac{1870121}{1260}\zeta_2^3
  - \mfrac{122474347}{1944}\zeta_5
  - \mfrac{116700197}{2916}\zeta_3 \zeta_2
  - \mfrac{31578173}{14580}\zeta_2^2
  - \mfrac{1935211669}{52488}\zeta_3
  + \mfrac{338298976}{6561}\zeta_2
  - \mfrac{4866548501}{236196}
 \Big)\bigg]
 \notag\\ &
+ \CS{C_F C_A^3} \bigg[
 \pole{\frac{1}{\ep^5}} \Big(
  - \mfrac{1331}{216}
 \Big)+ \pole{\frac{1}{\ep^4}} \Big(
   \mfrac{121}{12}\zeta_2
  - \mfrac{1067}{24}
 \Big)+ \pole{\frac{1}{\ep^3}} \Big(
  - \mfrac{121}{10}\zeta_2^2
  + \mfrac{3025}{12}\zeta_3
  - \mfrac{1927}{108}\zeta_2
  - \mfrac{39727}{216}
 \Big)+ \pole{\frac{1}{\ep^2}} \Big(
   \mfrac{1}{2}\zeta_3^2
  + \mfrac{626}{105}\zeta_2^3
 \notag\\ &
  - \mfrac{13013}{36}\zeta_5
  - \mint{77}\zeta_3 \zeta_2
  + \mfrac{2893}{12}\zeta_2^2
  + \mfrac{164719}{81}\zeta_3
  - \mfrac{14963}{36}\zeta_2
  - \mfrac{4881089}{11664}
 \Big)+ \pole{\frac{1}{\ep}} \Big(
   \mfrac{45511}{48}\zeta_7
  - \mfrac{206}{3}\zeta_5 \zeta_2
  + \mfrac{1033}{30}\zeta_3 \zeta_2^2
  - \mfrac{52547}{36}\zeta_3^2
 \notag\\ &
  - \mfrac{44204}{135}\zeta_2^3
  + \mfrac{136685}{108}\zeta_5
  - \mfrac{1936}{9}\zeta_3 \zeta_2
  + \mfrac{176134}{135}\zeta_2^2
  + \mfrac{573931}{54}\zeta_3
  - \mfrac{1980005}{648}\zeta_2
  + \mfrac{6220123}{5832}
 \Big)+  \Big(
  - \mfrac{14161}{30}\zeta_{5,3}
  + \mfrac{21577}{6}\zeta_5 \zeta_3
 \notag\\ &
  - \mfrac{1963}{3}\zeta_3^2 \zeta_2
  + \mfrac{10233079}{15750}\zeta_2^4
  + \mfrac{79877}{48}\zeta_7
  - \mfrac{7744}{9}\zeta_5 \zeta_2
  - \mfrac{71776}{45}\zeta_3 \zeta_2^2
  - \mfrac{2569009}{108}\zeta_3^2
  - \mfrac{8387389}{11340}\zeta_2^3
  + \mfrac{56929927}{3240}\zeta_5
  + \mfrac{2563075}{324}\zeta_3 \zeta_2
 \notag\\ &
  + \mfrac{802607}{180}\zeta_2^2
  + \mfrac{6792233}{162}\zeta_3
  - \mfrac{208858685}{11664}\zeta_2
  + \mfrac{3047508671}{139968}
 \Big)\bigg]
+ \CS{\frac{d^{abcd}_F d^{abcd}_A}{N_F}} \bigg[
 \pole{\frac{1}{\ep^2}} \Big(
   \mint{12}\zeta_3^2
  + \mfrac{248}{35}\zeta_2^3
  - \mfrac{110}{3}\zeta_5
  - \mfrac{4}{3}\zeta_3
  + \mint{4}\zeta_2
 \Big)
  \notag\\ &
  + \pole{\frac{1}{\ep}} \Big(
  - \mfrac{871}{2}\zeta_7
  - \mint{128}\zeta_5 \zeta_2
  + \mfrac{92}{5}\zeta_3 \zeta_2^2
  + \mfrac{418}{3}\zeta_3^2
  - \mfrac{3476}{315}\zeta_2^3
  + \mfrac{230}{9}\zeta_5
  + \mint{224}\zeta_3 \zeta_2
  - \mfrac{28}{15}\zeta_2^2
  + \mfrac{1516}{9}\zeta_3
  + \mfrac{272}{3}\zeta_2
  - \mint{32}
 \Big)+  \Big(
   \mint{260}\zeta_{5,3}
 \notag\\ &
  - \mint{5092}\zeta_5 \zeta_3
  - \mint{16}\zeta_3^2 \zeta_2
  - \mfrac{496766}{525}\zeta_2^4
  - \mint{1228}\zeta_7
  - \mfrac{12808}{3}\zeta_5 \zeta_2
  + \mfrac{14216}{15}\zeta_3 \zeta_2^2
  + \mfrac{72674}{9}\zeta_3^2
  + \mfrac{768632}{945}\zeta_2^3
  - \mfrac{65546}{27}\zeta_5
  + \mfrac{2516}{3}\zeta_3 \zeta_2
 \notag\\ &
  + \mfrac{8692}{45}\zeta_2^2
  + \mfrac{112346}{27}\zeta_3
  + \mfrac{8194}{9}\zeta_2
  - \mfrac{1588}{3}
 \Big)\bigg]
+ \CS{n_f C_F^3} \bigg[
 \pole{\frac{1}{\ep^7}} \Big(
   \mfrac{2}{3}
 \Big)+ \pole{\frac{1}{\ep^6}} \Big(
   \mfrac{10}{9}
 \Big)+ \pole{\frac{1}{\ep^5}} \Big(
   \mfrac{116}{27}
 \Big)+ \pole{\frac{1}{\ep^4}} \Big(
  - \mfrac{92}{3}\zeta_3
  + \mfrac{38}{3}\zeta_2
  + \mfrac{898}{81}
 \Big)
 \notag\\ &
 + \pole{\frac{1}{\ep^3}} \Big(
  - \mfrac{844}{45}\zeta_2^2
  - \mfrac{1318}{9}\zeta_3
  + \mfrac{178}{3}\zeta_2
  + \mfrac{67063}{1944}
 \Big)+ \pole{\frac{1}{\ep^2}} \Big(
  - \mfrac{4234}{45}\zeta_5
  + \mfrac{736}{9}\zeta_3 \zeta_2
  - \mfrac{5318}{135}\zeta_2^2
  - \mfrac{38501}{54}\zeta_3
  + \mfrac{2207}{9}\zeta_2
  + \mfrac{586997}{5832}
 \Big)
 \notag\\ &
 + \pole{\frac{1}{\ep}} \Big(
   \mfrac{36410}{27}\zeta_3^2
  + \mfrac{169268}{945}\zeta_2^3
  - \mfrac{55498}{27}\zeta_5
  - \mfrac{6784}{27}\zeta_3 \zeta_2
  - \mfrac{97709}{810}\zeta_2^2
  - \mfrac{199787}{81}\zeta_3
  + \mfrac{114713}{108}\zeta_2
  + \mfrac{5808187}{69984}
 \Big)+  \Big(
   \mfrac{1153615}{126}\zeta_7
 \notag\\ &
  + \mfrac{6316}{9}\zeta_5 \zeta_2
  + \mfrac{229468}{135}\zeta_3 \zeta_2^2
  + \mfrac{124198}{81}\zeta_3^2
  - \mfrac{342292}{567}\zeta_2^3
  - \mfrac{13555259}{810}\zeta_5
  - \mfrac{223877}{81}\zeta_3 \zeta_2
  + \mfrac{222727}{1215}\zeta_2^2
  - \mfrac{639475}{972}\zeta_3
  + \mfrac{405517}{81}\zeta_2
 \notag\\ &
  - \mfrac{252681119}{104976}
 \Big)\bigg]
+ \CS{n_f C_F^2 C_A} \bigg[
 \pole{\frac{1}{\ep^6}} \Big(
  - \mfrac{451}{162}
 \Big)+ \pole{\frac{1}{\ep^5}} \Big(
   \mfrac{41}{27}\zeta_2
  - \mfrac{2681}{243}
 \Big)+ \pole{\frac{1}{\ep^4}} \Big(
   \mfrac{629}{27}\zeta_3
  - \mfrac{598}{81}\zeta_2
  - \mfrac{17945}{486}
 \Big)+ \pole{\frac{1}{\ep^3}} \Big(
   \mfrac{2737}{135}\zeta_2^2
  + \mfrac{56570}{243}\zeta_3
 \notag\\ &
  - \mfrac{503}{6}\zeta_2
  - \mfrac{1887077}{17496}
 \Big)+ \pole{\frac{1}{\ep^2}} \Big(
   \mfrac{632}{3}\zeta_5
  - \mfrac{7328}{81}\zeta_3 \zeta_2
  + \mfrac{50002}{405}\zeta_2^2
  + \mfrac{1051067}{729}\zeta_3
  - \mfrac{1532465}{2916}\zeta_2
  - \mfrac{28195033}{104976}
 \Big)+ \pole{\frac{1}{\ep}} \Big(
  - \mfrac{81152}{81}\zeta_3^2
 \notag\\ &
  + \mfrac{8954}{315}\zeta_2^3
  + \mfrac{607772}{405}\zeta_5
  + \mfrac{27844}{243}\zeta_3 \zeta_2
  + \mfrac{188048}{405}\zeta_2^2
  + \mfrac{20440703}{2916}\zeta_3
  - \mfrac{46910953}{17496}\zeta_2
  - \mfrac{11426783}{23328}
 \Big)+  \Big(
   \mfrac{5669}{4}\zeta_7
  - \mfrac{249194}{135}\zeta_5 \zeta_2
 \notag\\ &
  - \mfrac{417244}{405}\zeta_3 \zeta_2^2
  - \mfrac{8430847}{729}\zeta_3^2
  - \mfrac{41398}{45}\zeta_2^3
  + \mfrac{46976113}{2430}\zeta_5
  + \mfrac{132347}{27}\zeta_3 \zeta_2
  + \mfrac{9178327}{7290}\zeta_2^2
  + \mfrac{1524037079}{52488}\zeta_3
  - \mfrac{447302531}{34992}\zeta_2
 \notag\\ &
  + \mfrac{635981455}{3779136}
 \Big)\bigg]
+ \CS{n_f C_F C_A^2} \bigg[
 \pole{\frac{1}{\ep^5}} \Big(
   \mfrac{121}{36}
 \Big)+ \pole{\frac{1}{\ep^4}} \Big(
  - \mfrac{11}{3}\zeta_2
  + \mfrac{2435}{108}
 \Big)+ \pole{\frac{1}{\ep^3}} \Big(
   \mfrac{11}{5}\zeta_2^2
  - \mfrac{242}{3}\zeta_3
  + \mfrac{173}{9}\zeta_2
  + \mfrac{121619}{1296}
 \Big)+ \pole{\frac{1}{\ep^2}} \Big(
   \mfrac{1093}{18}\zeta_5
 \notag\\ &
  + \mint{10}\zeta_3 \zeta_2
  - \mfrac{1313}{15}\zeta_2^2
  - \mfrac{19375}{27}\zeta_3
  + \mfrac{11839}{54}\zeta_2
  + \mfrac{2078443}{7776}
 \Big)+ \pole{\frac{1}{\ep}} \Big(
   \mfrac{2815}{18}\zeta_3^2
  + \mfrac{7493}{135}\zeta_2^3
  - \mfrac{20639}{27}\zeta_5
  + \mfrac{2540}{9}\zeta_3 \zeta_2
  - \mfrac{16729}{45}\zeta_2^2
 \notag\\ &
  - \mfrac{679399}{162}\zeta_3
  + \mfrac{872417}{648}\zeta_2
  + \mfrac{7711783}{46656}
 \Big)+  \Big(
   \mfrac{6943}{24}\zeta_7
  + \mfrac{2755}{9}\zeta_5 \zeta_2
  + \mfrac{1912}{45}\zeta_3 \zeta_2^2
  + \mfrac{182662}{27}\zeta_3^2
  + \mfrac{1798639}{11340}\zeta_2^3
  - \mfrac{29781037}{3240}\zeta_5
 \notag\\ &
  - \mfrac{115961}{108}\zeta_3 \zeta_2
  - \mfrac{728249}{540}\zeta_2^2
  - \mfrac{40929727}{1944}\zeta_3
  + \mfrac{27793927}{3888}\zeta_2
  - \mfrac{1377145789}{279936}
 \Big)\bigg]
+ \CS{n_f \frac{d^{abcd}_F d^{abcd}_F}{N_F}} \bigg[
 \pole{\frac{1}{\ep^2}} \Big(
   \mfrac{40}{3}\zeta_5
  + \mfrac{8}{3}\zeta_3
  - \mint{8}\zeta_2
 \Big)
 \notag\\ &
 + \pole{\frac{1}{\ep}} \Big(
  - \mfrac{152}{3}\zeta_3^2
  - \mfrac{1184}{315}\zeta_2^3
  + \mfrac{2720}{9}\zeta_5
  - \mint{16}\zeta_3 \zeta_2
  + \mfrac{40}{3}\zeta_2^2
  - \mfrac{416}{9}\zeta_3
  - \mfrac{568}{3}\zeta_2
  + \mint{64}
 \Big)+  \Big(
  - \mint{1240}\zeta_7
  + \mfrac{992}{3}\zeta_5 \zeta_2
  - \mfrac{3952}{15}\zeta_3 \zeta_2^2
 \notag\\ &
  - \mfrac{4504}{9}\zeta_3^2
  + \mfrac{215876}{945}\zeta_2^3
  + \mfrac{101938}{27}\zeta_5
  + \mfrac{572}{3}\zeta_3 \zeta_2
  - \mfrac{8}{45}\zeta_2^2
  - \mfrac{18202}{27}\zeta_3
  - \mfrac{18254}{9}\zeta_2
  + \mfrac{3488}{3}
 \Big)\bigg]
+ \CS{n_f^2 C_F^2} \bigg[
 \pole{\frac{1}{\ep^6}} \Big(
   \mfrac{41}{162}
 \Big)+ \pole{\frac{1}{\ep^5}} \Big(
   \mfrac{205}{243}
 \Big)
 \notag\\ &
 + \pole{\frac{1}{\ep^4}} \Big(
   \mfrac{73}{81}\zeta_2
  + \mfrac{497}{162}
 \Big)+ \pole{\frac{1}{\ep^3}} \Big(
  - \mfrac{2620}{243}\zeta_3
  + \mfrac{1702}{243}\zeta_2
  + \mfrac{86477}{8748}
 \Big)+ \pole{\frac{1}{\ep^2}} \Big(
  - \mfrac{1072}{405}\zeta_2^2
  - \mfrac{58276}{729}\zeta_3
  + \mfrac{3091}{81}\zeta_2
  + \mfrac{209642}{6561}
 \Big)+ \pole{\frac{1}{\ep}} \Big(
   \mfrac{8948}{405}\zeta_5
 \notag\\ &
  + \mfrac{7808}{243}\zeta_3 \zeta_2
  - \mfrac{11692}{1215}\zeta_2^2
  - \mfrac{123817}{243}\zeta_3
  + \mfrac{752095}{4374}\zeta_2
  + \mfrac{11881193}{104976}
 \Big)+  \Big(
   \mfrac{689582}{729}\zeta_3^2
  + \mfrac{191252}{945}\zeta_2^3
  - \mfrac{217004}{243}\zeta_5
  + \mfrac{60584}{729}\zeta_3 \zeta_2
 \notag\\ &
  - \mfrac{41431}{405}\zeta_2^2
  - \mfrac{42764887}{13122}\zeta_3
  + \mfrac{9851863}{13122}\zeta_2
  + \mfrac{436438747}{944784}
 \Big)\bigg]
+ \CS{n_f^2 C_F C_A} \bigg[
 \pole{\frac{1}{\ep^5}} \Big(
  - \mfrac{11}{18}
 \Big)+ \pole{\frac{1}{\ep^4}} \Big(
   \mfrac{1}{3}\zeta_2
  - \mfrac{197}{54}
 \Big)+ \pole{\frac{1}{\ep^3}} \Big(
   \mfrac{19}{3}\zeta_3
  - \mint{5}\zeta_2
 \notag\\ &
  - \mfrac{19819}{1296}
 \Big)+ \pole{\frac{1}{\ep^2}} \Big(
   \mfrac{39}{5}\zeta_2^2
  + \mfrac{1846}{27}\zeta_3
  - \mfrac{982}{27}\zeta_2
  - \mfrac{417697}{7776}
 \Big)+ \pole{\frac{1}{\ep}} \Big(
   \mfrac{784}{9}\zeta_5
  - \mfrac{412}{9}\zeta_3 \zeta_2
  + \mfrac{502}{45}\zeta_2^2
  + \mfrac{35089}{81}\zeta_3
  - \mfrac{121495}{648}\zeta_2
  - \mfrac{7185425}{46656}
 \Big)
 \notag\\ &
 +  \Big(
  - \mfrac{1714}{3}\zeta_3^2
  - \mfrac{2836}{315}\zeta_2^3
  + \mfrac{144946}{135}\zeta_5
  + \mfrac{28}{3}\zeta_3 \zeta_2
  + \mfrac{4258}{135}\zeta_2^2
  + \mfrac{1209401}{486}\zeta_3
  - \mfrac{1168375}{1296}\zeta_2
  - \mfrac{73152481}{279936}
 \Big)\bigg]
+ \CS{n_f^3 C_F} \bigg[
 \pole{\frac{1}{\ep^5}} \Big(
   \mfrac{1}{27}
 \Big)
 \notag\\ &
 + \pole{\frac{1}{\ep^4}} \Big(
   \mfrac{5}{27}
 \Big)+ \pole{\frac{1}{\ep^3}} \Big(
   \mfrac{10}{27}\zeta_2
  + \mfrac{65}{81}
 \Big)+ \pole{\frac{1}{\ep^2}} \Big(
  - \mfrac{82}{81}\zeta_3
  + \mfrac{50}{27}\zeta_2
  + \mfrac{2537}{729}
 \Big)+ \pole{\frac{1}{\ep}} \Big(
   \mfrac{302}{135}\zeta_2^2
  - \mfrac{410}{81}\zeta_3
  + \mfrac{650}{81}\zeta_2
  + \mfrac{11408}{729}
 \Big)+  \Big(
  - \mfrac{2194}{135}\zeta_5
 \notag\\ &
  - \mfrac{820}{81}\zeta_3 \zeta_2
  + \mfrac{302}{27}\zeta_2^2
  - \mfrac{5330}{243}\zeta_3
  + \mfrac{25370}{729}\zeta_2
  + \mfrac{159908}{2187}
 \Big)\bigg]
 + \mathcal{O}(\pole{\ep^1})

%% file: eq/beta.tex
\beta_0 &=
\CS{C_A}
\Big(
   \mfrac{11}{3}
\Big)
+ \CS{n_f}
\Big(
   - \mfrac{2}{3}
\Big)
,
\\
\beta_1 &=
\CS{C_A^2}
\Big(
   \mfrac{34}{3}
\Big)
+ \CS{n_f C_A}
\Big(
   - \mfrac{10}{3}
\Big)
+ \CS{n_f C_F}
\Big(
   - \mint{2}
\Big)
,
\\
\beta_2 &=
\CS{C_A^3}
\Big(
   \mfrac{2857}{54}
 \Big)
+ \CS{n_f C_A^2}
\Big(
  - \mfrac{1415}{54}
 \Big)
+ \CS{n_f C_A C_F}
\Big(
  - \mfrac{205}{18}
 \Big)
+ \CS{n_f C_F^2}
\Big(
   \mint{1}
 \Big)
+ \CS{n_f^2 C_A}
\Big(
   \mfrac{79}{54}
 \Big)
+ \CS{n_f^2 C_F}
\Big(
   \mfrac{11}{9}
 \Big),
\\
\beta_3 &=
\CS{C_A^4}
\Big(
  - \mfrac{44}{9}\zeta_3
  + \mfrac{150653}{486}
 \Big)
+ \CS{\frac{d^{abcd}_A d^{abcd}_A}{N_A}}
\Big(
   \mfrac{704}{3}\zeta_3
  - \mfrac{80}{9}
 \Big)
+ \CS{n_f C_A^3}
\Big(
   \mfrac{68}{3}\zeta_3
  - \mfrac{39143}{162}
 \Big)
+ \CS{n_f C_A^2 C_F}
\Big(
  - \mfrac{328}{9}\zeta_3
  + \mfrac{7073}{486}
 \Big)
 \notag \\ &
+ \CS{n_f C_A C_F^2}
\Big(
   \mfrac{176}{9}\zeta_3
  - \mfrac{2102}{27}
 \Big)
+ \CS{n_f C_F^3}
\Big(
   \mint{23}
 \Big)
+ \CS{n_f \frac{d^{abcd}_A d^{abcd}_F}{N_A}}
\Big(
  - \mfrac{1664}{3}\zeta_3
  + \mfrac{512}{9}
 \Big)
+ \CS{n_f^2 C_A^2}
\Big(
   \mfrac{56}{9}\zeta_3
  + \mfrac{3965}{162}
 \Big)
 \notag \\ &
+ \CS{n_f^2 C_A C_F}
\Big(
   \mfrac{112}{9}\zeta_3
  + \mfrac{4288}{243}
 \Big)
+ \CS{n_f^2 C_F^2}
\Big(
  - \mfrac{176}{9}\zeta_3
  + \mfrac{338}{27}
 \Big)
+ \CS{n_f^2 \frac{d^{abcd}_F d^{abcd}_F}{N_A}}
\Big(
   \mfrac{512}{3}\zeta_3
  - \mfrac{704}{9}
 \Big)
+ \CS{n_f^3 C_A}
\Big(
   \mfrac{53}{243}
 \Big)
 \notag \\ &
+ \CS{n_f^3 C_F}
\Big(
   \mfrac{154}{243}
 \Big)
,

%% file: eq/mass.tex
\gamma^m_0 &=
\CS{C_F} \big(
  \mint{3}
\big)
,\\
\gamma^m_1 &=
\CS{C_F C_A} \Big(
  \mfrac{97}{6}
\Big)
+ \CS{C_F^2}
\Big(
  \mfrac{3}{2}
\Big)
+ \CS{n_f C_F}
\Big(
  - \mfrac{10}{6}
\Big)
,\\
\gamma^m_2 &=
\CS{C_F^3}
\Big(
  \mfrac{129}{2}
 \Big)
+ \CS{C_F^2 C_A}
\Big(
  - \mfrac{129}{4}
 \Big)
+ \CS{C_F C_A^2}
\Big(
   \mfrac{11413}{108}
 \Big)
+ \CS{n_f C_F^2}
\Big(
   \mint{24}\zeta_3
  - \mint{23}
 \Big)
+ \CS{n_f C_F C_A}
\Big(
  - \mint{24}\zeta_3
  - \mfrac{278}{27}
 \Big)
 \notag \\ &
+ \CS{n_f^2 C_F}
\Big(
  - \mfrac{35}{27}
 \Big),
\\
\gamma^m_3 &=
\CS{C_F^4}
\Big(
  - \mint{336}\zeta_3
  - \mfrac{1261}{8}
 \Big)
+ \CS{C_F^3 C_A}
\Big(
   \mint{316}\zeta_3
  + \mfrac{15349}{12}
 \Big)
+ \CS{C_F^2 C_A^2}
\Big(
   \mint{440}\zeta_5
  - \mint{152}\zeta_3
  - \mfrac{34045}{36}
 \Big)
+ \CS{C_F C_A^3}
\Big(
  - \mint{440}\zeta_5
  + \mfrac{1418}{9}\zeta_3
 \notag \\ &
  + \mfrac{70055}{72}
 \Big)
+ \CS{\frac{d^{abcd}_F d^{abcd}_A}{N_F}}
\Big(
   \mint{240}\zeta_3
  - \mint{32}
 \Big)
+ \CS{n_f C_F^3}
\Big(
  - \mint{240}\zeta_5
  + \mint{276}\zeta_3
  - \mfrac{140}{3}
 \Big)
+ \CS{n_f C_F^2 C_A}
\Big(
  - \mint{132}\zeta_4
  + \mint{40}\zeta_5
  + \mint{184}\zeta_3
 \notag \\ &
  - \mfrac{8819}{54}
 \Big)
+ \CS{n_f C_F C_A^2}
\Big(
   \mint{132}\zeta_4
  + \mint{200}\zeta_5
  - \mfrac{1342}{3}\zeta_3
  - \mfrac{65459}{324}
 \Big)
+ \CS{n_f \frac{d^{abcd}_F d^{abcd}_F}{N_F}}
\Big(
  - \mint{480}\zeta_3
  + \mint{64}
 \Big)
+ \CS{n_f^2 C_F^2}
\Big(
   \mint{24}\zeta_4
  - \mint{40}\zeta_3
  + \mfrac{76}{27}
 \Big)
 \notag \\ &
+ \CS{n_f^2 C_F C_A}
\Big(
  - \mint{24}\zeta_4
  + \mint{40}\zeta_3
  + \mfrac{671}{162}
 \Big)
+ \CS{n_f^3 C_F}
\Big(
   \mfrac{16}{9}\zeta_3
  - \mfrac{83}{81}
 \Big)

%% file: eq/cusp.tex
\Gamma^r_1 &= \CS{C_R}\Big( \mint{4} \Big),
\\
\Gamma^r_2 &= \CS{C_R C_A}\Big( - \mint{8}\zeta_2 + \mfrac{268}{9} \Big)
+ \CS{n_f C_R} \Big(- \mfrac{40}{9}\Big),
\\
\Gamma^r_3 &= \CS{C_R C_A^2} \Big(
   \mfrac{176}{5}\zeta_2^2
  + \mfrac{88}{3}\zeta_3
  - \mfrac{1072}{9}\zeta_2
  + \mfrac{490}{3}
 \Big)
+ \CS{n_f C_R C_A} \Big(
  - \mfrac{112}{3}\zeta_3
  + \mfrac{160}{9}\zeta_2
  - \mfrac{836}{27}
 \Big)
+ \CS{n_f C_R C_F} \Big(
   \mint{32}\zeta_3
  - \mfrac{110}{3}
 \Big)
 \notag \\ &
+ \CS{n_f^2 C_R} \Big(
  - \mfrac{16}{27}
 \Big),
\\
\Gamma^r_4 &= 
\CS{C_R C_A^3} \Big(
    -\mint{16} \zeta _3^2
    -\mfrac{20032}{105} \zeta _2^3
    -\mfrac{3608}{9} \zeta _5
    -\mfrac{352}{3} \zeta _3 \zeta _2
    +\mfrac{3608}{5} \zeta _2^2
    +\mfrac{20944}{27} \zeta _3
    -\mfrac{88400}{81} \zeta _2
    +\mfrac{84278}{81}
\Big)
+ \CS{\frac{d^{abcd}_R d^{abcd}_A}{N_R}} \Big(
    -384 \zeta _3^2
\nonumber\\&
    -\mfrac{7936}{35} \zeta _2^3
    +\mfrac{3520}{3} \zeta _5
    +\mfrac{128}{3} \zeta _3
    -\mint{128} \zeta _2
\Big)
+ \CS{n_f C_R C_A^2} \Big(
    \mfrac{2096}{9} \zeta _5
    +\mfrac{448}{3} \zeta _3 \zeta _2
    -\mfrac{352}{15} \zeta _2^2
    -\mfrac{23104}{27} \zeta _3
    +\mfrac{20320}{81} \zeta _2
    -\mfrac{24137}{81}
\Big)
\nonumber\\&
+\CS{n_f C_R C_A C_F} \Big(
    \mint{160} \zeta _5
    -\mint{128} \zeta _3 \zeta _2
    -\mfrac{352}{5} \zeta _2^2 
    +\mfrac{3712}{9} \zeta _3
    +\mfrac{440}{3} \zeta _2
    -\mfrac{34066}{81}
\Big) 
+ \CS{n_f C_R C_F^2}  \Big(
    -\mint{320} \zeta _5
    +\mfrac{592}{3} \zeta _3
    +\mfrac{572}{9}
\Big)
\nonumber\\&
+ \CS{n_f \frac{d^{abcd}_R d^{abcd}_F}{N_R}} \Big(
    -\mfrac{1280}{3} \zeta _5
    -\mfrac{256}{3} \zeta _3
    +\mint{256} \zeta _2
\Big)
+ \CS{n_f^2 C_R C_A} \Big(
    -\mfrac{224}{15} \zeta _2^2
    +\mfrac{2240}{27} \zeta _3
    -\mfrac{608}{81} \zeta _2
    +\mfrac{923}{81}
\Big) 
+ \CS{n_f^2 C_R C_F} \Big(
    \mfrac{64}{5} \zeta _2^2
\nonumber \\ &
    -\mfrac{640}{9} \zeta _3
    +\mfrac{2392}{81}\Big)
+ \CS{n_f^3 C_R} \Big(
    \mfrac{64}{27} \zeta _3
    -\mfrac{32}{81}
\Big)

%% file: eq/colq.tex
\gamma_1^q &= \CS{C_F} \Big( \mint{6} \Big),
\\
\gamma_2^q &=
 \CS{C_F^2}\Big( \mint{48} \zeta_3 - \mint{24} \zeta_2 + \mint{3}\Big)
+ \CS{C_F C_A}\Big(- \mint{52} \zeta_3  + \mint{22} \zeta_2 + \mfrac{961}{27}\Big)
+ \CS{n_f C_F}\Big( - \mint{4} \zeta_2 -\mfrac{130}{27}\Big),
\\
\gamma_3^q &= \CS{C_F^3} \Big(
  - \mint{480}\zeta_5
  - \mint{64}\zeta_3 \zeta_2
  + \mfrac{576}{5}\zeta_2^2
  + \mint{136}\zeta_3
  + \mint{36}\zeta_2
  + \mint{29}
 \Big)
+ \CS{C_F^2 C_A} \Big(
   \mint{240}\zeta_5
  + \mint{32}\zeta_3 \zeta_2
  - \mfrac{1976}{15}\zeta_2^2
  + \mfrac{1688}{3}\zeta_3
  - \mfrac{820}{3}\zeta_2
  + \mfrac{151}{2}
 \Big)
 \notag \\ &
+ \CS{C_F C_A^2} \Big(
   \mint{272}\zeta_5
  + \mfrac{176}{3}\zeta_3 \zeta_2
  + \mfrac{332}{5}\zeta_2^2
  - \mfrac{7052}{9}\zeta_3
  + \mfrac{14326}{81}\zeta_2
  + \mfrac{139345}{1458}
 \Big)
+ \CS{n_f C_F^2} \Big(
   \mfrac{112}{3}\zeta_2^2
  - \mfrac{512}{9}\zeta_3
  + \mfrac{52}{3}\zeta_2
  - \mfrac{2953}{27}
 \Big)
 \notag \\ &
+ \CS{n_f C_F C_A} \Big(
  - \mfrac{88}{5}\zeta_2^2
  + \mfrac{1928}{27}\zeta_3
  - \mfrac{5188}{81}\zeta_2
  + \mfrac{17318}{729}
 \Big)
+ \CS{n_f^2 C_F} \Big(
   \mfrac{16}{27}\zeta_3
  + \mfrac{40}{9}\zeta_2
  - \mfrac{4834}{729}
 \Big),
\\
\gamma_4^q &=
\CS{C_F^4}
\Big(
   \mint{11760}\zeta_7
  - \mint{768}\zeta_5 \zeta_2
  + \mfrac{256}{5}\zeta_3 \zeta_2^2
  - \mint{2304}\zeta_3^2
  - \mfrac{33776}{35}\zeta_2^3
  - \mint{5040}\zeta_5
  - \mint{240}\zeta_3 \zeta_2
  - \mfrac{1368}{5}\zeta_2^2
  + \mint{4008}\zeta_3
  - \mint{900}\zeta_2
  + \mfrac{4873}{12}
 \Big)
 \notag \\ &
+ \CS{C_F^3 C_A}
\Big(
  - \mint{21840}\zeta_7
  + \mint{4128}\zeta_5 \zeta_2
  + \mfrac{512}{5}\zeta_3 \zeta_2^2
  + \mint{6440}\zeta_3^2
  + \mfrac{634376}{315}\zeta_2^3
  - \mint{1952}\zeta_5
  - \mfrac{3976}{3}\zeta_3 \zeta_2
  + \mfrac{8668}{5}\zeta_2^2
  - \mint{6520}\zeta_3
  + \mint{2334}\zeta_2
 \notag \\ &
  - \mfrac{2085}{2}
 \Big)
+ \CS{C_F^2 C_A^2}
\Big(
   \mint{17220}\zeta_7
  - \mint{4208}\zeta_5 \zeta_2
  - \mfrac{128}{5}\zeta_3 \zeta_2^2
  - \mfrac{14204}{3}\zeta_3^2
  - \mfrac{43976}{35}\zeta_2^3
  + \mfrac{10708}{9}\zeta_5
  + \mfrac{4192}{9}\zeta_3 \zeta_2
  - \mfrac{48680}{27}\zeta_2^2
  + \mfrac{259324}{27}\zeta_3
 \notag \\ &
  - \mfrac{93542}{27}\zeta_2
  + \mfrac{29639}{18}
 \Big)
+ \CS{C_F C_A^3}
\Big(
  - \mfrac{45511}{6}\zeta_7
  + \mfrac{1648}{3}\zeta_5 \zeta_2
  - \mfrac{4132}{15}\zeta_3 \zeta_2^2
  + \mfrac{5126}{9}\zeta_3^2
  - \mfrac{77152}{315}\zeta_2^3
  + \mfrac{175166}{27}\zeta_5
  + \mfrac{15400}{9}\zeta_3 \zeta_2
 \notag \\ &
  + \mfrac{186742}{135}\zeta_2^2
  - \mfrac{1751224}{243}\zeta_3
  + \mfrac{1062149}{729}\zeta_2
  + \mfrac{7179083}{26244}
 \Big)
+ \CS{\frac{d^{abcd}_F d^{abcd}_A}{N_F}}
\Big(
   \mint{3484}\zeta_7
  + \mint{1024}\zeta_5 \zeta_2
  - \mfrac{736}{5}\zeta_3 \zeta_2^2
  - \mfrac{3344}{3}\zeta_3^2
  + \mfrac{27808}{315}\zeta_2^3
 \notag \\ &
  - \mfrac{1840}{9}\zeta_5
  - \mint{1792}\zeta_3 \zeta_2
  + \mfrac{224}{15}\zeta_2^2
  - \mfrac{7808}{9}\zeta_3
  - \mfrac{2176}{3}\zeta_2
  + \mint{192}
 \Big)
+ \CS{n_f C_F^3}
\Big(
   \mint{368}\zeta_3^2
  - \mfrac{117344}{315}\zeta_2^3
  + \mfrac{3872}{3}\zeta_5
  - \mfrac{512}{3}\zeta_3 \zeta_2
  - \mfrac{668}{5}\zeta_2^2
 \notag \\ &
  - \mfrac{1120}{9}\zeta_3
  + \mint{322}\zeta_2
  + \mfrac{27949}{108}
 \Big)
+ \CS{n_f C_F^2 C_A}
\Big(
  - \mfrac{3400}{3}\zeta_3^2
  + \mfrac{5744}{35}\zeta_2^3
  - \mfrac{4472}{3}\zeta_5
  + \mfrac{3904}{9}\zeta_3 \zeta_2
  + \mfrac{105488}{135}\zeta_2^2
  - \mfrac{23518}{81}\zeta_3
  + \mfrac{673}{27}\zeta_2
 \notag \\ &
  - \mfrac{1092511}{972}
 \Big)
+ \CS{n_f C_F C_A^2}
\Big(
   \mfrac{6916}{9}\zeta_3^2
  + \mfrac{24184}{315}\zeta_2^3
  + \mfrac{6088}{27}\zeta_5
  - \mfrac{3584}{9}\zeta_3 \zeta_2
  - \mfrac{17164}{45}\zeta_2^2
  + \mfrac{140632}{243}\zeta_3
  - \mfrac{445117}{729}\zeta_2
  + \mfrac{326863}{1944}
 \Big)
 \notag \\ &
+ \CS{n_f \frac{d^{abcd}_F d^{abcd}_F}{N_F}}
\Big(
   \mfrac{1216}{3}\zeta_3^2
  + \mfrac{9472}{315}\zeta_2^3
  - \mfrac{21760}{9}\zeta_5
  + \mint{128}\zeta_3 \zeta_2
  - \mfrac{320}{3}\zeta_2^2
  - \mfrac{5312}{9}\zeta_3
  + \mfrac{4544}{3}\zeta_2
  - \mint{384}
 \Big)
+ \CS{n_f^2 C_F^2}
\Big(
   \mfrac{1040}{9}\zeta_5
 \notag \\ &
  - \mfrac{224}{9}\zeta_3 \zeta_2
  - \mfrac{8032}{135}\zeta_2^2
  - \mfrac{4232}{81}\zeta_3
  + \mfrac{1972}{27}\zeta_2
  + \mfrac{9965}{486}
 \Big)
+ \CS{n_f^2 C_F C_A}
\Big(
  - \mfrac{1184}{9}\zeta_5
  + \mfrac{256}{9}\zeta_3 \zeta_2
  + \mfrac{152}{15}\zeta_2^2
  + \mfrac{14872}{243}\zeta_3
  + \mfrac{41579}{729}\zeta_2
 \notag \\ &
  - \mfrac{97189}{17496}
 \Big)
+ \CS{n_f^3 C_F} \Big(
   \mfrac{128}{135}\zeta_2^2
  + \mfrac{1424}{243}\zeta_3
  + \mfrac{16}{27}\zeta_2
  - \mfrac{37382}{6561}
 \Big)

%% file: eq/colg.tex
\gamma^g_1 &= \CS{C_A}
  \Big(
   \mfrac{22}{3}
 \Big)
+ \CS{n_f}
  \Big(
  - \mfrac{4}{3}
 \Big)
,
\\
\gamma^g_2 &=
\CS{C_A^2}
  \Big(
  - \mint{4}\zeta_3
  - \mfrac{22}{3}\zeta_2
  + \mfrac{1384}{27}
 \Big)
+ \CS{n_f C_A}
  \Big(
   \mfrac{4}{3}\zeta_2
  - \mfrac{256}{27}
 \Big)
+ \CS{n_f C_F}
  \Big(
  - \mint{4}
 \Big)
 ,
\\
\gamma^g_3 &= \CS{C_A^3}
  \Big(
   \mint{32}\zeta_5
  + \mfrac{80}{3}\zeta_3 \zeta_2
  + \mfrac{1276}{15}\zeta_2^2
  - \mfrac{244}{3}\zeta_3
  - \mfrac{12218}{81}\zeta_2
  + \mfrac{194372}{729}
 \Big)
+ \CS{n_f C_A^2}
  \Big(
  - \mfrac{328}{15}\zeta_2^2
  - \mfrac{712}{27}\zeta_3
  + \mfrac{2396}{81}\zeta_2
  - \mfrac{30715}{729}
 \Big)
 \notag \\ &
+ \CS{n_f C_A C_F}
  \Big(
   \mfrac{32}{5}\zeta_2^2
  + \mfrac{304}{9}\zeta_3
  + \mint{4}\zeta_2
  - \mfrac{2434}{27}
 \Big)
+ \CS{n_f C_F^2}
  \Big(
   \mint{2}
 \Big)
+ \CS{n_f^2 C_A}
  \Big(
   \mfrac{112}{27}\zeta_3
  - \mfrac{40}{27}\zeta_2
  + \mfrac{269}{729}
 \Big)
+ \CS{n_f^2 C_F}
  \Big(
   \mfrac{22}{9}
 \Big)
 ,
\\
\gamma^g_4 &=
 \CS{C_A^4}
  \Big(
  - \mfrac{2671}{6}\zeta_7
  - \mfrac{896}{3}\zeta_5 \zeta_2
  - \mfrac{2212}{15}\zeta_3 \zeta_2^2
  - \mfrac{286}{9}\zeta_3^2
  - \mfrac{674696}{945}\zeta_2^3
  + \mfrac{19232}{27}\zeta_5
  + \mfrac{1588}{3}\zeta_3 \zeta_2
  + \mfrac{249448}{135}\zeta_2^2
  + \mfrac{36380}{243}\zeta_3
  - \mfrac{1051411}{729}\zeta_2
 \notag \\ &
  + \mfrac{10672040}{6561}
 \Big)
+ \CS{\frac{d_ {abcd}^A d_ {abcd}^A}{N_A}}
  \Big(
   \mint{3484}\zeta_7
  + \mint{1024}\zeta_5 \zeta_2
  - \mfrac{736}{5}\zeta_3 \zeta_2^2
  - \mfrac{3344}{3}\zeta_3^2
  + \mfrac{39776}{315}\zeta_2^3
  + \mfrac{2720}{9}\zeta_5
  - \mint{2336}\zeta_3 \zeta_2
  - \mfrac{1808}{15}\zeta_2^2
 \notag \\ &
  - \mfrac{12512}{9}\zeta_3
  + \mint{64}\zeta_2
  + \mfrac{128}{9}
 \Big)
+ \CS{n_f C_A^3}
  \Big(
  - \mfrac{596}{9}\zeta_3^2
  + \mfrac{148976}{945}\zeta_2^3
  + \mfrac{16066}{27}\zeta_5
  + \mint{148}\zeta_3 \zeta_2
  - \mfrac{69502}{135}\zeta_2^2
  - \mfrac{260822}{243}\zeta_3
  + \mfrac{155273}{729}\zeta_2
 \notag \\ &
  - \mfrac{421325}{1944}
 \Big)
+ \CS{n_f C_A^2 C_F}
  \Big(
   \mint{152}\zeta_3^2
  + \mfrac{5632}{315}\zeta_2^3
  + \mfrac{8}{9}\zeta_5
  - \mint{176}\zeta_3 \zeta_2
  - \mfrac{1196}{45}\zeta_2^2
  + \mfrac{29606}{81}\zeta_3
  + \mfrac{3023}{9}\zeta_2
  - \mfrac{903983}{972}
 \Big)
 \notag \\ &
+ \CS{n_f C_A C_F^2}
  \Big(
  - \mint{80}\zeta_3^2
  - \mfrac{320}{7}\zeta_2^3
  - \mfrac{1600}{3}\zeta_5
  + \mfrac{148}{5}\zeta_2^2
  + \mfrac{1592}{3}\zeta_3
  - \mint{2}\zeta_2
  + \mfrac{685}{12}
 \Big)
+ \CS{n_f C_F^3}
  \Big(
   \mint{46}
 \Big)
+ \CS{n_f \frac{d_ {abcd}^A d_ {abcd}^F}{N_A}}
  \Big(
   \mfrac{1216}{3}\zeta_3^2
 \notag \\ &
  - \mfrac{14464}{315}\zeta_2^3
  - \mfrac{30880}{9}\zeta_5
  + \mint{1216}\zeta_3 \zeta_2
  + \mfrac{2464}{15}\zeta_2^2
  + \mfrac{2560}{9}\zeta_3
  - \mint{64}\zeta_2
  + \mfrac{448}{9}
 \Big)
+ \CS{n_f^2 C_A^2}
  \Big(
  - \mfrac{1024}{9}\zeta_5
  - \mint{32}\zeta_3 \zeta_2
  + \mfrac{3128}{135}\zeta_2^2
  + \mfrac{37354}{243}\zeta_3
 \notag \\ &
  - \mfrac{13483}{729}\zeta_2
  + \mfrac{611939}{17496}
 \Big)
+ \CS{n_f^2 C_A C_F}
  \Big(
   \mfrac{304}{9}\zeta_5
  + \mfrac{32}{3}\zeta_3 \zeta_2
  + \mfrac{128}{45}\zeta_2^2
  - \mfrac{1688}{81}\zeta_3
  - \mfrac{172}{9}\zeta_2
  + \mfrac{1199}{18}
 \Big)
+ \CS{n_f^2 C_F^2}
  \Big(
  - \mfrac{352}{9}\zeta_3
  + \mfrac{676}{27}
 \Big)
 \notag \\ &
+ \CS{n_f^2 \frac{d_{abcd}^F d_{abcd}^F}{N_A}}
  \Big(
   \mfrac{1024}{3}\zeta_3
  - \mfrac{1408}{9}
 \Big)
+ \CS{n_f^3 C_A}
  \Big(
   \mfrac{256}{135}\zeta_2^2
  - \mfrac{400}{243}\zeta_3
  - \mfrac{16}{81}\zeta_2
  - \mfrac{15890}{6561}
 \Big)
+ \CS{n_f^3 C_F}
  \Big(
   \mfrac{308}{243}
 \Big)

%% file: eq/cb.tex
C^b_1 &= \CS{C_F} \bigg[\logpow{L}^{2}\big(-1\big) + \big(\zeta_2 - \mint{2}\big)\bigg] \, ,\\[1.0em]
C^b_2 &= \CS{C_F^2} \bigg[\logpow{L^4} \Big(\mfrac{1}{2}\Big)+\logpow{L^2} \Big(-\zeta_2+2\Big)+\logpow{L} \Big(24 \zeta_3-12 \zeta_2\Big)+\Big(-\mfrac{83}{10}  \zeta_2^2-30 \zeta_3+14 \zeta_2+6\Big)\bigg]
+\CS{C_F C_A}\bigg[ \logpow{L^3} \Big(\mfrac{11}{9}\Big)+\logpow{L^2} \Big(2 \zeta_2
\notag\\ &
-\mfrac{67}{9}\Big)
+\logpow{L} \Big(-26 \zeta_3+\mfrac{22}{3} \zeta_2+\mfrac{242}{27}\Big)+\Big(\mfrac{44}{5}  \zeta_2^2+\mfrac{151}{9} \zeta_3-\mfrac{103}{18}  \zeta_2-\mfrac{467}{81}\Big)\bigg]
+\CS{C_F n_f} \bigg[\logpow{L^3}\Big(-\mfrac{2}{9}\Big)+\logpow{L^2}\Big(\mfrac{10}{9}\Big)+\logpow{L} \Big(-\mfrac{4}{3} \zeta_2
\notag\\ &
-\mfrac{56}{27}\Big)
+\Big(\mfrac{5}{9} \zeta_2+\mfrac{2}{9} \zeta_3+\mfrac{200}{81}\Big)\bigg]\, ,\\[1.0em]
C^b_3 &=
\CS{C_F^3} \bigg[
 \logpow{L^{6}} \Big(
  - \mfrac{1}{6}
 \Big)+ \logpow{L^{4}} \Big(
   \mfrac{1}{2}\zeta_2
  - \mint{1}
 \Big)+ \logpow{L^{3}} \Big(
  - \mint{24}\zeta_3
  + \mint{12}\zeta_2
 \Big)+ \logpow{L^{2}} \Big(
   \mfrac{83}{10}\zeta_2^2
  + \mint{30}\zeta_3
  - \mint{14}\zeta_2
  - \mint{6}
 \Big)+ \logpow{L} \Big(
  - \mint{240}\zeta_5
  - \mint{8}\zeta_3 \zeta_2
 \notag \\ &
  + \mfrac{228}{5}\zeta_2^2
  + \mint{20}\zeta_3
  + \mint{42}\zeta_2
  - \mint{50}
 \Big)+ \logpow{} \Big(
   \mint{16}\zeta_3^2
  + \mfrac{37729}{630}\zeta_2^3
  + \mint{424}\zeta_5
  + \mint{178}\zeta_3 \zeta_2
  - \mint{77}\zeta_2^2
  - \mint{654}\zeta_3
  - \mfrac{353}{3}\zeta_2
  + \mfrac{575}{3}
 \Big)
 \bigg]
 \notag \\ &
+ \CS{C_F^2 C_A} \bigg[
 \logpow{L^{5}} \Big(
  - \mfrac{11}{9}
 \Big)+ \logpow{L^{4}} \Big(
  - \mint{2}\zeta_2
  + \mfrac{67}{9}
 \Big)+ \logpow{L^{3}} \Big(
   \mint{26}\zeta_3
  - \mfrac{55}{9}\zeta_2
  - \mfrac{308}{27}
 \Big)+ \logpow{L^{2}} \Big(
  - \mfrac{34}{5}\zeta_2^2
  - \mfrac{943}{9}\zeta_3
  + \mfrac{689}{18}\zeta_2
  + \mfrac{1673}{81}
 \Big)
 \notag \\ &
 + \logpow{L} \Big(
   \mint{120}\zeta_5
  - \mint{10}\zeta_3 \zeta_2
  + \mint{6}\zeta_2^2
  + \mfrac{1660}{3}\zeta_3
  - \mfrac{7012}{27}\zeta_2
  + \mfrac{614}{27}
 \Big)+ \logpow{} \Big(
   \mfrac{296}{3}\zeta_3^2
  - \mfrac{12676}{315}\zeta_2^3
  - \mfrac{1676}{9}\zeta_5
  - \mfrac{3049}{9}\zeta_3 \zeta_2
  - \mfrac{893}{270}\zeta_2^2
  - \mfrac{4820}{27}\zeta_3
 \notag \\ &
  + \mfrac{31819}{81}\zeta_2
  - \mfrac{9335}{81}
 \Big)
 \bigg]
+ \CS{C_F C_A^2} \bigg[
 \logpow{L^{4}} \Big(
  - \mfrac{121}{54}
 \Big)+ \logpow{L^{3}} \Big(
  - \mfrac{44}{9}\zeta_2
  + \mfrac{1780}{81}
 \Big)+ \logpow{L^{2}} \Big(
  - \mfrac{44}{5}\zeta_2^2
  + \mint{88}\zeta_3
  + \mfrac{26}{9}\zeta_2
  - \mfrac{11939}{162}
 \Big)+ \logpow{L} \Big(
   \mint{136}\zeta_5
 \notag \\ &
  + \mfrac{88}{3}\zeta_3 \zeta_2
  - \mfrac{94}{3}\zeta_2^2
  - \mfrac{13900}{27}\zeta_3
  + \mfrac{9644}{81}\zeta_2
  + \mfrac{10289}{1458}
 \Big)+ \logpow{} \Big(
  - \mfrac{1136}{9}\zeta_3^2
  - \mfrac{6152}{189}\zeta_2^3
  + \mfrac{106}{9}\zeta_5
  + \mfrac{326}{3}\zeta_3 \zeta_2
  + \mfrac{10093}{135}\zeta_2^2
  + \mfrac{107648}{243}\zeta_3
 \notag \\ &
  - \mfrac{264515}{1458}\zeta_2
  + \mfrac{5964431}{26244}
 \Big)
 \bigg]
+ \CS{n_f C_F^2} \bigg[
 \logpow{L^{5}} \Big(
   \mfrac{2}{9}
 \Big)+ \logpow{L^{4}} \Big(
  - \mfrac{10}{9}
 \Big)+ \logpow{L^{3}} \Big(
   \mfrac{10}{9}\zeta_2
  + \mfrac{50}{27}
 \Big)+ \logpow{L^{2}} \Big(
   \mfrac{70}{9}\zeta_3
  - \mfrac{67}{9}\zeta_2
  + \mfrac{725}{162}
 \Big)+ \logpow{L} \Big(
   \mfrac{28}{5}\zeta_2^2
 \notag \\ &
  - \mfrac{832}{9}\zeta_3
  + \mfrac{880}{27}\zeta_2
  - \mfrac{1415}{54}
 \Big)+ \logpow{} \Big(
  - \mfrac{416}{9}\zeta_5
  - \mfrac{38}{9}\zeta_3 \zeta_2
  - \mfrac{61}{27}\zeta_2^2
  + \mfrac{11996}{81}\zeta_3
  - \mfrac{6131}{162}\zeta_2
  + \mfrac{35875}{972}
 \Big)
 \bigg]
+ \CS{n_f C_F C_A} \bigg[
 \logpow{L^{4}} \Big(
   \mfrac{22}{27}
 \Big)
 \notag \\ &
 + \logpow{L^{3}} \Big(
   \mfrac{8}{9}\zeta_2
  - \mfrac{578}{81}
 \Big)+ \logpow{L^{2}} \Big(
  - \mint{8}\zeta_3
  + \mfrac{16}{3}\zeta_2
  + \mfrac{1727}{81}
 \Big)+ \logpow{L} \Big(
   \mfrac{44}{15}\zeta_2^2
  + \mfrac{724}{9}\zeta_3
  - \mfrac{3272}{81}\zeta_2
  - \mfrac{7499}{729}
 \Big)+ \logpow{} \Big(
  - \mfrac{4}{3}\zeta_5
  + \mfrac{4}{3}\zeta_3 \zeta_2
  - \mfrac{476}{135}\zeta_2^2
 \notag \\ &
  - \mfrac{2860}{27}\zeta_3
  + \mfrac{33259}{729}\zeta_2
  - \mfrac{521975}{13122}
 \Big)
 \bigg]
+ \CS{n_f^2 C_F} \bigg[
 \logpow{L^{4}} \Big(
  - \mfrac{2}{27}
 \Big)+ \logpow{L^{3}} \Big(
   \mfrac{40}{81}
 \Big)+ \logpow{L^{2}} \Big(
  - \mfrac{8}{9}\zeta_2
  - \mfrac{100}{81}
 \Big)+ \logpow{L} \Big(
   \mfrac{16}{27}\zeta_3
  + \mfrac{80}{27}\zeta_2
  + \mfrac{928}{729}
 \Big)
 \notag \\ &
 + \logpow{} \Big(
  - \mfrac{188}{135}\zeta_2^2
  - \mfrac{200}{243}\zeta_3
  - \mfrac{212}{81}\zeta_2
  + \mfrac{2072}{6561}
 \Big)
 \bigg]
\, ,\\[1.0em]
C^b_4 &=
\CS{C_F^4} \bigg[
 \logpow{L^{8}} \Big(
   \mfrac{1}{24}
 \Big)+ \logpow{L^{6}} \Big(
  - \mfrac{1}{6}\zeta_2
  + \mfrac{1}{3}
 \Big)+ \logpow{L^{5}} \Big(
   \mint{12}\zeta_3
  - \mint{6}\zeta_2
 \Big)+ \logpow{L^{4}} \Big(
  - \mfrac{83}{20}\zeta_2^2
  - \mint{15}\zeta_3
  + \mint{7}\zeta_2
  + \mint{3}
 \Big)+ \logpow{L^{3}} \Big(
   \mint{240}\zeta_5
  + \mint{8}\zeta_3 \zeta_2
  - \mfrac{228}{5}\zeta_2^2
\notag\\ &
  - \mint{20}\zeta_3
  - \mint{42}\zeta_2
  + \mint{50}
 \Big)+ \logpow{L^{2}} \Big(
   \mint{272}\zeta_3^2
  - \mfrac{37729}{630}\zeta_2^3
  - \mint{424}\zeta_5
  - \mint{466}\zeta_3 \zeta_2
  + \mint{149}\zeta_2^2
  + \mint{654}\zeta_3
  + \mfrac{353}{3}\zeta_2
  - \mfrac{575}{3}
 \Big)+ \logpow{L} \Big(
   \mint{5880}\zeta_7
\notag\\ &
  - \mint{624}\zeta_5 \zeta_2
  - \mfrac{1028}{5}\zeta_3 \zeta_2^2
  - \mint{1872}\zeta_3^2
  - \mfrac{11386}{35}\zeta_2^3
  - \mint{2040}\zeta_5
  + \mint{708}\zeta_3 \zeta_2
  - \mint{402}\zeta_2^2
  + \mint{2348}\zeta_3
  - \mint{608}\zeta_2
  + \mfrac{1382}{3}
 \Big)
 + \logpow{} \Big(
  - \mfrac{2208}{5}\zeta_{5,3}
\notag\\ &
  - \mint{1792}\zeta_5 \zeta_3
  + \mint{840}\zeta_3^2 \zeta_2
  - \mfrac{7508687}{63000}\zeta_2^4
  - \mfrac{12321}{2}\zeta_7
  - \mint{4448}\zeta_5 \zeta_2
  + \mfrac{2081}{5}\zeta_3 \zeta_2^2
  + \mint{2940}\zeta_3^2
  + \mfrac{31403}{45}\zeta_2^3
  + \mint{13323}\zeta_5
  + \mint{292}\zeta_3 \zeta_2
\notag\\ &
  + \mfrac{972}{5}\zeta_2^2
  - \mint{9275}\zeta_3
  + \mfrac{7029}{4}\zeta_2
  - \mfrac{22259}{12}
 \Big)
 \bigg]
+ \CS{C_F^3 C_A} \bigg[
 \logpow{L^{7}} \Big(
   \mfrac{11}{18}
 \Big)+ \logpow{L^{6}} \Big(
   \zeta_2
  - \mfrac{67}{18}
 \Big)+ \logpow{L^{5}} \Big(
  - \mint{13}\zeta_3
  + \mfrac{22}{9}\zeta_2
  + \mfrac{187}{27}
 \Big)+ \logpow{L^{4}} \Big(
   \mfrac{12}{5}\zeta_2^2
\notag\\ &
  + \mfrac{2263}{18}\zeta_3
  - \mfrac{601}{12}\zeta_2
  - \mfrac{2879}{162}
 \Big)+ \logpow{L^{3}} \Big(
  - \mint{120}\zeta_5
  + \mint{58}\zeta_3 \zeta_2
  - \mfrac{3613}{90}\zeta_2^2
  - \mfrac{2306}{3}\zeta_3
  + \mfrac{9886}{27}\zeta_2
  - \mfrac{416}{27}
 \Big)+ \logpow{L^{2}} \Big(
  - \mfrac{2168}{3}\zeta_3^2
  + \mfrac{7447}{315}\zeta_2^3
\notag\\ &
  + \mfrac{13556}{9}\zeta_5
  + \mfrac{7693}{9}\zeta_3 \zeta_2
  - \mfrac{7228}{27}\zeta_2^2
  + \mfrac{11312}{27}\zeta_3
  - \mfrac{63148}{81}\zeta_2
  + \mfrac{27992}{81}
 \Big)+ \logpow{L} \Big(
  - \mint{10920}\zeta_7
  + \mint{2184}\zeta_5 \zeta_2
  + \mfrac{2471}{5}\zeta_3 \zeta_2^2
  + \mfrac{12680}{3}\zeta_3^2
\notag\\ &
  + \mfrac{8969}{105}\zeta_2^3
  - \mint{5880}\zeta_5
  - \mint{3404}\zeta_3 \zeta_2
  + \mfrac{258931}{135}\zeta_2^2
  + \mfrac{73226}{27}\zeta_3
  + \mfrac{8888}{3}\zeta_2
  - \mfrac{36350}{9}
 \Big)
 + \logpow{} \Big(
  - \mfrac{692}{5}\zeta_{5,3}
  + \mint{3696}\zeta_5 \zeta_3
  - \mfrac{8536}{3}\zeta_3^2 \zeta_2
\notag\\ &
  + \mfrac{506012}{1125}\zeta_2^4
  + \mfrac{178357}{24}\zeta_7
  + \mfrac{83443}{9}\zeta_5 \zeta_2
  - \mfrac{107401}{90}\zeta_3 \zeta_2^2
  - \mint{2880}\zeta_3^2
  + \mfrac{1145267}{3780}\zeta_2^3
  - \mfrac{73607}{36}\zeta_5
  + \mfrac{320363}{54}\zeta_3 \zeta_2
  - \mfrac{3878479}{1620}\zeta_2^2
\notag\\ &
  - \mfrac{526531}{36}\zeta_3
  - \mfrac{4901615}{648}\zeta_2
  + \mfrac{2888701}{216}
 \Big)
 \bigg]
+ \CS{C_F^2 C_A^2} \bigg[
 \logpow{L^{6}} \Big(
   \mfrac{242}{81}
 \Big)+ \logpow{L^{5}} \Big(
   \mfrac{22}{3}\zeta_2
  - \mfrac{839}{27}
 \Big)+ \logpow{L^{4}} \Big(
   \mfrac{54}{5}\zeta_2^2
  - \mfrac{1078}{9}\zeta_3
  - \mfrac{199}{18}\zeta_2
  + \mfrac{28393}{243}
 \Big)
\notag\\ &
 + \logpow{L^{3}} \Big(
  - \mint{136}\zeta_5
  - \mfrac{244}{3}\zeta_3 \zeta_2
  + \mfrac{778}{15}\zeta_2^2
  + \mfrac{85175}{81}\zeta_3
  - \mfrac{47353}{162}\zeta_2
  - \mfrac{181927}{1458}
 \Big)+ \logpow{L^{2}} \Big(
   \mfrac{4178}{9}\zeta_3^2
  + \mfrac{39076}{945}\zeta_2^3
  - \mfrac{6046}{9}\zeta_5
  - \mfrac{2392}{9}\zeta_3 \zeta_2
\notag\\ &
  - \mfrac{13096}{135}\zeta_2^2
  - \mfrac{973523}{243}\zeta_3
  + \mfrac{2524883}{1458}\zeta_2
  - \mfrac{5785547}{26244}
 \Big)+ \logpow{L} \Big(
   \mint{8610}\zeta_7
  - \mint{1968}\zeta_5 \zeta_2
  - \mfrac{3184}{15}\zeta_3 \zeta_2^2
  - \mfrac{35000}{9}\zeta_3^2
  - \mfrac{17506}{315}\zeta_2^3
  + \mfrac{5794}{3}\zeta_5
\notag\\ &
  + \mfrac{11527}{3}\zeta_3 \zeta_2
  - \mfrac{284933}{405}\zeta_2^2
  + \mfrac{2079788}{243}\zeta_3
  - \mfrac{9736801}{1458}\zeta_2
  + \mfrac{6576527}{2187}
 \Big)
 + \logpow{} \Big(
   \mint{1046}\zeta_{5,3}
  - \mint{5104}\zeta_5 \zeta_3
  + \mfrac{24208}{9}\zeta_3^2 \zeta_2
  - \mfrac{3829877}{4725}\zeta_2^4
\notag\\ &
  - \mfrac{105405}{16}\zeta_7
  - \mfrac{91561}{18}\zeta_5 \zeta_2
  + \mfrac{64541}{45}\zeta_3 \zeta_2^2
  + \mfrac{697187}{81}\zeta_3^2
  + \mfrac{113683}{1260}\zeta_2^3
  - \mfrac{125555}{216}\zeta_5
  - \mfrac{12580021}{972}\zeta_3 \zeta_2
  + \mfrac{52786259}{29160}\zeta_2^2
  + \mfrac{29217731}{5832}\zeta_3
\notag\\ &
  + \mfrac{279041783}{26244}\zeta_2
  - \mfrac{526960807}{52488}
 \Big)
 \bigg]
+ \CS{C_F C_A^3} \bigg[
 \logpow{L^{5}} \Big(
   \mfrac{1331}{270}
 \Big)+ \logpow{L^{4}} \Big(
   \mfrac{121}{9}\zeta_2
  - \mfrac{5456}{81}
 \Big)+ \logpow{L^{3}} \Big(
   \mfrac{484}{15}\zeta_2^2
  - \mfrac{968}{3}\zeta_3
  - \mfrac{694}{27}\zeta_2
  + \mfrac{83618}{243}
 \Big)
\notag\\ &
 + \logpow{L^{2}} \Big(
   \mint{4}\zeta_3^2
  + \mfrac{5008}{105}\zeta_2^3
  - \mfrac{5830}{9}\zeta_5
  - \mint{132}\zeta_3 \zeta_2
  - \mfrac{121}{15}\zeta_2^2
  + \mfrac{26390}{9}\zeta_3
  - \mfrac{4186}{9}\zeta_2
  - \mfrac{1167889}{2916}
 \Big)+ \logpow{L} \Big(
  - \mfrac{45511}{12}\zeta_7
  + \mfrac{824}{3}\zeta_5 \zeta_2
\notag\\ &
  - \mfrac{2066}{15}\zeta_3 \zeta_2^2
  + \mfrac{15059}{9}\zeta_3^2
  + \mfrac{222632}{945}\zeta_2^3
  + \mfrac{95965}{27}\zeta_5
  - \mfrac{3058}{9}\zeta_3 \zeta_2
  - \mfrac{2972}{9}\zeta_2^2
  - \mfrac{243382}{27}\zeta_3
  + \mfrac{1361261}{486}\zeta_2
  - \mfrac{27115865}{8748}
 \Big)
\notag\\ &
 + \logpow{} \Big(
  - \mfrac{14161}{30}\zeta_{5,3}
  + \mfrac{21577}{6}\zeta_5 \zeta_3
  - \mfrac{1963}{3}\zeta_3^2 \zeta_2
  + \mfrac{10233079}{15750}\zeta_2^4
  + \mfrac{258199}{144}\zeta_7
  + \mint{1056}\zeta_5 \zeta_2
  - \mfrac{23288}{45}\zeta_3 \zeta_2^2
  - \mfrac{702221}{108}\zeta_3^2
  - \mfrac{2000759}{2268}\zeta_2^3
\notag\\ &
  - \mfrac{9786737}{3240}\zeta_5
  + \mfrac{444085}{108}\zeta_3 \zeta_2
  + \mfrac{184637}{810}\zeta_2^2
  + \mfrac{8121343}{1458}\zeta_3
  - \mfrac{146447531}{34992}\zeta_2
  + \mfrac{3966128773}{419904}
 \Big)
 \bigg]
+ \CS{\frac{d^{abcd}_F d^{abcd}_A}{N_F}} \bigg[
 \logpow{L^{2}} \Big(
   \mint{96}\zeta_3^2
  + \mfrac{1984}{35}\zeta_2^3
\notag\\ &
  - \mfrac{880}{3}\zeta_5
  - \mfrac{32}{3}\zeta_3
  + \mint{32}\zeta_2
 \Big)+ \logpow{L} \Big(
   \mint{1742}\zeta_7
  + \mint{512}\zeta_5 \zeta_2
  - \mfrac{368}{5}\zeta_3 \zeta_2^2
  - \mfrac{1672}{3}\zeta_3^2
  + \mfrac{13904}{315}\zeta_2^3
  - \mfrac{920}{9}\zeta_5
  - \mint{896}\zeta_3 \zeta_2
  + \mfrac{112}{15}\zeta_2^2
  - \mfrac{6064}{9}\zeta_3
\notag\\ &
  - \mfrac{1088}{3}\zeta_2
  + \mint{128}
 \Big)
 + \logpow{} \Big(
   \mint{260}\zeta_{5,3}
  - \mint{5092}\zeta_5 \zeta_3
  - \mint{16}\zeta_3^2 \zeta_2
  - \mfrac{496766}{525}\zeta_2^4
  - \mint{1228}\zeta_7
  - \mfrac{12808}{3}\zeta_5 \zeta_2
  + \mfrac{14216}{15}\zeta_3 \zeta_2^2
  + \mfrac{72674}{9}\zeta_3^2
\notag\\ &
  + \mfrac{768632}{945}\zeta_2^3
  - \mfrac{65546}{27}\zeta_5
  + \mfrac{2516}{3}\zeta_3 \zeta_2
  + \mfrac{8692}{45}\zeta_2^2
  + \mfrac{112346}{27}\zeta_3
  + \mfrac{8194}{9}\zeta_2
  - \mfrac{1588}{3}
 \Big)
 \bigg]
+ \CS{n_f C_F^3} \bigg[
 \logpow{L^{7}} \Big(
  - \mfrac{1}{9}
 \Big)+ \logpow{L^{6}} \Big(
   \mfrac{5}{9}
 \Big)
\notag\\ &
 + \logpow{L^{5}} \Big(
  - \mfrac{4}{9}\zeta_2
  - \mfrac{22}{27}
 \Big)+ \logpow{L^{4}} \Big(
  - \mfrac{119}{9}\zeta_3
  + \mfrac{59}{6}\zeta_2
  - \mfrac{925}{162}
 \Big)+ \logpow{L^{3}} \Big(
  - \mfrac{169}{45}\zeta_2^2
  + \mfrac{1132}{9}\zeta_3
  - \mfrac{1342}{27}\zeta_2
  + \mfrac{1433}{54}
 \Big)+ \logpow{L^{2}} \Big(
  - \mfrac{1024}{9}\zeta_5
\notag\\ &
  - \mfrac{466}{9}\zeta_3 \zeta_2
  + \mfrac{7916}{135}\zeta_2^2
  - \mfrac{14624}{81}\zeta_3
  + \mfrac{7894}{81}\zeta_2
  - \mfrac{111259}{972}
 \Big)+ \logpow{L} \Big(
   \mfrac{664}{3}\zeta_3^2
  - \mfrac{3278}{105}\zeta_2^3
  + \mfrac{5200}{3}\zeta_5
  + \mfrac{2600}{9}\zeta_3 \zeta_2
  - \mfrac{8402}{27}\zeta_2^2
  - \mfrac{42656}{27}\zeta_3
\notag\\ &
  - \mfrac{6499}{54}\zeta_2
  + \mfrac{46391}{72}
 \Big)
 + \logpow{} \Big(
   \mfrac{2013}{2}\zeta_7
  - \mfrac{1124}{9}\zeta_5 \zeta_2
  - \mfrac{7567}{45}\zeta_3 \zeta_2^2
  - \mfrac{3764}{3}\zeta_3^2
  - \mfrac{107227}{1890}\zeta_2^3
  - \mfrac{70907}{18}\zeta_5
  - \mfrac{72811}{81}\zeta_3 \zeta_2
  + \mfrac{432143}{810}\zeta_2^2
\notag\\ &
  + \mfrac{1934375}{324}\zeta_3
  + \mfrac{172627}{972}\zeta_2
  - \mfrac{6554087}{3888}
 \Big)
 \bigg]
+ \CS{n_f C_F^2 C_A} \bigg[
 \logpow{L^{6}} \Big(
  - \mfrac{88}{81}
 \Big)+ \logpow{L^{5}} \Big(
  - \mfrac{4}{3}\zeta_2
  + \mfrac{274}{27}
 \Big)+ \logpow{L^{4}} \Big(
   \mfrac{124}{9}\zeta_3
  - \mfrac{50}{9}\zeta_2
  - \mfrac{15889}{486}
 \Big)
\notag\\ &
 + \logpow{L^{3}} \Big(
  - \mfrac{20}{3}\zeta_2^2
  - \mfrac{16264}{81}\zeta_3
  + \mfrac{8804}{81}\zeta_2
  + \mfrac{5047}{729}
 \Big)+ \logpow{L^{2}} \Big(
   \mfrac{244}{3}\zeta_5
  + \mfrac{664}{9}\zeta_3 \zeta_2
  - \mfrac{1606}{135}\zeta_2^2
  + \mfrac{91328}{81}\zeta_3
  - \mfrac{408010}{729}\zeta_2
  + \mfrac{7464559}{26244}
 \Big)
\notag\\ &
 + \logpow{L} \Big(
  - \mfrac{3376}{9}\zeta_3^2
  - \mfrac{1564}{63}\zeta_2^3
  - \mfrac{1948}{3}\zeta_5
  - \mfrac{10802}{27}\zeta_3 \zeta_2
  + \mfrac{177476}{405}\zeta_2^2
  - \mfrac{642457}{243}\zeta_3
  + \mfrac{2079563}{1458}\zeta_2
  - \mfrac{18844181}{17496}
 \Big)
 + \logpow{} \Big(
  - \mfrac{1219}{4}\zeta_7
\notag\\ &
  + \mint{114}\zeta_5 \zeta_2
  + \mfrac{15934}{45}\zeta_3 \zeta_2^2
  + \mfrac{9446}{81}\zeta_3^2
  - \mfrac{1846}{21}\zeta_2^3
  + \mfrac{45995}{18}\zeta_5
  + \mfrac{155563}{81}\zeta_3 \zeta_2
  - \mfrac{3347782}{3645}\zeta_2^2
  - \mfrac{1262017}{5832}\zeta_3
  - \mfrac{145213765}{104976}\zeta_2
\notag\\ &
  + \mfrac{756958495}{419904}
 \Big)
 \bigg]
+ \CS{n_f C_F C_A^2} \bigg[
 \logpow{L^{5}} \Big(
  - \mfrac{121}{45}
 \Big)+ \logpow{L^{4}} \Big(
  - \mfrac{44}{9}\zeta_2
  + \mfrac{1831}{54}
 \Big)+ \logpow{L^{3}} \Big(
  - \mfrac{88}{15}\zeta_2^2
  + \mint{88}\zeta_3
  - \mfrac{356}{27}\zeta_2
  - \mfrac{8693}{54}
 \Big)+ \logpow{L^{2}} \Big(
   \mfrac{700}{9}\zeta_5
\notag\\ &
  - \mint{8}\zeta_3 \zeta_2
  - \mfrac{208}{5}\zeta_2^2
  - \mint{830}\zeta_3
  + \mfrac{8588}{27}\zeta_2
  + \mfrac{186151}{972}
 \Big)+ \logpow{L} \Big(
   \mfrac{1186}{9}\zeta_3^2
  - \mfrac{25244}{945}\zeta_2^3
  - \mfrac{1324}{27}\zeta_5
  + \mfrac{32}{9}\zeta_3 \zeta_2
  + \mfrac{52}{15}\zeta_2^2
  + \mfrac{234470}{81}\zeta_3
\notag\\ &
  - \mfrac{290270}{243}\zeta_2
  + \mfrac{12030185}{11664}
 \Big)
 + \logpow{} \Big(
   \mfrac{19141}{72}\zeta_7
  - \mfrac{127}{3}\zeta_5 \zeta_2
  - \mfrac{6904}{45}\zeta_3 \zeta_2^2
  + \mfrac{5462}{9}\zeta_3^2
  + \mfrac{153371}{2268}\zeta_2^3
  - \mfrac{2343853}{3240}\zeta_5
  - \mfrac{73985}{108}\zeta_3 \zeta_2
  + \mfrac{120913}{540}\zeta_2^2
\notag\\ &
  - \mfrac{16605365}{5832}\zeta_3
  + \mfrac{46423375}{34992}\zeta_2
  - \mfrac{2567430839}{839808}
 \Big)
 \bigg]
+ \CS{n_f \frac{d^{abcd}_F d^{abcd}_F}{N_F}} \bigg[
 \logpow{L^{2}} \Big(
   \mfrac{320}{3}\zeta_5
  + \mfrac{64}{3}\zeta_3
  - \mint{64}\zeta_2
 \Big)+ \logpow{L} \Big(
   \mfrac{608}{3}\zeta_3^2
  + \mfrac{4736}{315}\zeta_2^3
\notag\\ &
  - \mfrac{10880}{9}\zeta_5
  + \mint{64}\zeta_3 \zeta_2
  - \mfrac{160}{3}\zeta_2^2
  + \mfrac{1664}{9}\zeta_3
  + \mfrac{2272}{3}\zeta_2
  - \mint{256}
 \Big)
 + \logpow{} \Big(
  - \mint{1240}\zeta_7
  + \mfrac{992}{3}\zeta_5 \zeta_2
  - \mfrac{3952}{15}\zeta_3 \zeta_2^2
  - \mfrac{4504}{9}\zeta_3^2
  + \mfrac{215876}{945}\zeta_2^3
\notag\\ &
  + \mfrac{101938}{27}\zeta_5
  + \mfrac{572}{3}\zeta_3 \zeta_2
  - \mfrac{8}{45}\zeta_2^2
  - \mfrac{18202}{27}\zeta_3
  - \mfrac{18254}{9}\zeta_2
  + \mfrac{3488}{3}
 \Big)
 \bigg]
+ \CS{n_f^2 C_F^2} \bigg[
 \logpow{L^{6}} \Big(
   \mfrac{8}{81}
 \Big)+ \logpow{L^{5}} \Big(
  - \mfrac{20}{27}
 \Big)+ \logpow{L^{4}} \Big(
   \mfrac{10}{9}\zeta_2
  + \mfrac{463}{243}
 \Big)
\notag\\ &
 + \logpow{L^{3}} \Big(
   \mfrac{380}{81}\zeta_3
  - \mfrac{254}{27}\zeta_2
  + \mfrac{2104}{729}
 \Big)+ \logpow{L^{2}} \Big(
   \mfrac{692}{135}\zeta_2^2
  - \mfrac{17884}{243}\zeta_3
  + \mfrac{2906}{81}\zeta_2
  - \mfrac{457105}{13122}
 \Big)+ \logpow{L} \Big(
  - \mfrac{104}{3}\zeta_5
  - \mfrac{568}{27}\zeta_3 \zeta_2
  - \mfrac{1924}{45}\zeta_2^2
\notag\\ &
  + \mfrac{75380}{243}\zeta_3
  - \mfrac{36440}{729}\zeta_2
  + \mfrac{837923}{8748}
 \Big)
 + \logpow{} \Big(
   \mfrac{4556}{81}\zeta_3^2
  + \mfrac{3520}{189}\zeta_2^3
  - \mfrac{1568}{27}\zeta_5
  + \mfrac{3358}{243}\zeta_3 \zeta_2
  + \mfrac{45551}{810}\zeta_2^2
  - \mfrac{612127}{1458}\zeta_3
  + \mfrac{74333}{6561}\zeta_2
\notag\\ &
  - \mfrac{11290865}{104976}
 \Big)
 \bigg]
+ \CS{n_f^2 C_F C_A} \bigg[
 \logpow{L^{5}} \Big(
   \mfrac{22}{45}
 \Big)+ \logpow{L^{4}} \Big(
   \mfrac{4}{9}\zeta_2
  - \mfrac{143}{27}
 \Big)+ \logpow{L^{3}} \Big(
  - \mfrac{16}{3}\zeta_3
  + \mfrac{184}{27}\zeta_2
  + \mfrac{3515}{162}
 \Big)+ \logpow{L^{2}} \Big(
   \mfrac{20}{3}\zeta_2^2
  + \mfrac{508}{9}\zeta_3
  - \mfrac{1600}{27}\zeta_2
\notag\\ &
  - \mfrac{26293}{972}
 \Big)+ \logpow{L} \Big(
  - \mfrac{616}{9}\zeta_5
  + \mfrac{152}{9}\zeta_3 \zeta_2
  + \mfrac{344}{15}\zeta_2^2
  - \mfrac{17068}{81}\zeta_3
  + \mfrac{36643}{243}\zeta_2
  - \mfrac{823055}{11664}
 \Big)
 + \logpow{} \Big(
  - \mfrac{622}{27}\zeta_3^2
  + \mfrac{1654}{135}\zeta_2^3
  + \mfrac{19094}{135}\zeta_5
\notag\\ &
  - \mfrac{20}{27}\zeta_3 \zeta_2
  - \mfrac{1957}{27}\zeta_2^2
  + \mfrac{408781}{1458}\zeta_3
  - \mfrac{4264925}{34992}\zeta_2
  + \mfrac{176182813}{839808}
 \Big)
 \bigg]
+ \CS{n_f^3 C_F} \bigg[
 \logpow{L^{5}} \Big(
  - \mfrac{4}{135}
 \Big)+ \logpow{L^{4}} \Big(
   \mfrac{20}{81}
 \Big)+ \logpow{L^{3}} \Big(
  - \mfrac{16}{27}\zeta_2
  - \mfrac{200}{243}
 \Big)
\notag\\ &
 + \logpow{L^{2}} \Big(
   \mfrac{80}{27}\zeta_2
  + \mfrac{1000}{729}
 \Big)+ \logpow{L} \Big(
  - \mfrac{104}{45}\zeta_2^2
  - \mfrac{40}{81}\zeta_3
  - \mfrac{400}{81}\zeta_2
  - \mfrac{2608}{2187}
 \Big)
 + \logpow{} \Big(
  - \mfrac{106}{135}\zeta_5
  + \mfrac{4}{9}\zeta_3 \zeta_2
  + \mfrac{328}{81}\zeta_2^2
  + \mfrac{14}{243}\zeta_3
  + \mfrac{1946}{729}\zeta_2
\notag\\ &
  + \mfrac{6460}{6561}
 \Big)
 \bigg]

%% file: eq/cq.tex
C^q_4 &= \CS{C_F^4} \bigg[
 \logpow{L^{8}} \Big(
   \mfrac{1}{24}
 \Big)+ \logpow{L^{7}} \Big(
  - \mfrac{1}{2}
 \Big)+ \logpow{L^{6}} \Big(
  - \mfrac{1}{6}\zeta_2
  + \mfrac{43}{12}
 \Big)+ \logpow{L^{5}} \Big(
   \mint{12}\zeta_3
  - \mfrac{9}{2}\zeta_2
  - \mfrac{63}{4}
 \Big)+ \logpow{L^{4}} \Big(
  - \mfrac{83}{20}\zeta_2^2
  - \mint{87}\zeta_3
  + \mint{42}\zeta_2
  + \mfrac{813}{16}
 \Big)
 \notag\\ &
 + \logpow{L^{3}} \Big(
   \mint{240}\zeta_5
  + \mint{8}\zeta_3 \zeta_2
  - \mfrac{207}{10}\zeta_2^2
  + \mint{322}\zeta_3
  - \mint{228}\zeta_2
  - \mfrac{1019}{8}
 \Big)+ \logpow{L^{2}} \Big(
   \mint{272}\zeta_3^2
  - \mfrac{37729}{630}\zeta_2^3
  - \mint{1384}\zeta_5
  - \mint{562}\zeta_3 \zeta_2
  + \mfrac{5081}{20}\zeta_2^2
  - \zeta_3
  \notag\\ &
  + \mfrac{16603}{24}\zeta_2
  + \mfrac{17455}{48}
 \Big)+ \logpow{L} \Big(
   \mint{5880}\zeta_7
  - \mint{624}\zeta_5 \zeta_2
  - \mfrac{1028}{5}\zeta_3 \zeta_2^2
  - \mint{1824}\zeta_3^2
  - \mfrac{30587}{210}\zeta_2^3
  + \mint{1392}\zeta_5
  + \mint{1818}\zeta_3 \zeta_2
  - \mfrac{21837}{20}\zeta_2^2
  + \mint{770}\zeta_3
  \notag\\ &
  - \mfrac{13895}{8}\zeta_2
  - \mfrac{23995}{48}
 \Big) + \logpow{}\Big(
  - \mfrac{2208}{5}\zeta_{5,3}
  - \mint{1792}\zeta_5 \zeta_3
  + \mint{840}\zeta_3^2 \zeta_2
  - \mfrac{7508687}{63000}\zeta_2^4
  - \mfrac{29919}{2}\zeta_7
  - \mint{2696}\zeta_5 \zeta_2
  + \mfrac{2009}{5}\zeta_3 \zeta_2^2
  + \mint{5072}\zeta_3^2
  \notag\\ &
  + \mfrac{563503}{630}\zeta_2^3
  + \mfrac{44977}{3}\zeta_5
  - \mint{1930}\zeta_3 \zeta_2
  + \mfrac{19375}{16}\zeta_2^2
  - \mfrac{129505}{12}\zeta_3
  + \mfrac{26749}{8}\zeta_2
  + \mfrac{153365}{384}
 \Big)
 \bigg]
+ \CS{C_F^3 C_A} \bigg[
 \logpow{L^{7}} \Big(
   \mfrac{11}{18}
 \Big)+ \logpow{L^{6}} \Big(
   \zeta_2
  - \mfrac{365}{36}
 \Big)
  \notag \\ &
 + \logpow{L^{5}} \Big(
  - \mint{13}\zeta_3
  - \mfrac{32}{9}\zeta_2
  + \mfrac{8389}{108}
 \Big)+ \logpow{L^{4}} \Big(
   \mfrac{12}{5}\zeta_2^2
  + \mfrac{3829}{18}\zeta_3
  - \mfrac{581}{12}\zeta_2
  - \mfrac{472609}{1296}
 \Big)+ \logpow{L^{3}} \Big(
  - \mint{120}\zeta_5
  + \mint{58}\zeta_3 \zeta_2
  - \mfrac{5449}{90}\zeta_2^2
  - \mint{1542}\zeta_3
   \notag \\ &
  + \mfrac{36677}{54}\zeta_2
  + \mint{1101}
 \Big)+ \logpow{L^{2}} \Big(
  - \mfrac{2168}{3}\zeta_3^2
  + \mfrac{7447}{315}\zeta_2^3
  + \mfrac{17876}{9}\zeta_5
  + \mfrac{8125}{9}\zeta_3 \zeta_2
  - \mfrac{14629}{108}\zeta_2^2
  + \mfrac{258667}{54}\zeta_3
  - \mfrac{2084963}{648}\zeta_2
  - \mfrac{1547747}{648}
 \Big)
  \notag \\ &
 + \logpow{L} \Big(
  - \mint{10920}\zeta_7
  + \mint{2184}\zeta_5 \zeta_2
  + \mfrac{2471}{5}\zeta_3 \zeta_2^2
  + \mfrac{14864}{3}\zeta_3^2
  - \mfrac{3707}{105}\zeta_2^3
  - \mfrac{30476}{3}\zeta_5
  - \mfrac{18757}{3}\zeta_3 \zeta_2
  + \mfrac{1313227}{540}\zeta_2^2
  - \mfrac{608191}{108}\zeta_3
   \notag \\ &
  + \mfrac{676297}{72}\zeta_2
  + \mfrac{46493}{12}
 \Big) + \logpow{}\Big(
  - \mfrac{692}{5}\zeta_{5,3}
  + \mint{3696}\zeta_5 \zeta_3
  - \mfrac{8536}{3}\zeta_3^2 \zeta_2
  + \mfrac{506012}{1125}\zeta_2^4
  + \mfrac{474205}{24}\zeta_7
  + \mfrac{37975}{9}\zeta_5 \zeta_2
  - \mfrac{113287}{90}\zeta_3 \zeta_2^2
   \notag \\ &
  - \mint{8504}\zeta_3^2
  + \mfrac{2013857}{3780}\zeta_2^3
  + \mfrac{325717}{36}\zeta_5
  + \mfrac{787613}{54}\zeta_3 \zeta_2
  - \mfrac{32251333}{6480}\zeta_2^2
  - \mfrac{288281}{72}\zeta_3
  - \mfrac{6575143}{432}\zeta_2
  - \mfrac{1147289}{192}
 \Big)
 \bigg]
+ \CS{C_F^2 C_A^2} \bigg[
 \logpow{L^{6}} \Big(
   \mfrac{242}{81}
 \Big)
  \notag \\ &
 + \logpow{L^{5}} \Big(
   \mfrac{22}{3}\zeta_2
  - \mfrac{1565}{27}
 \Big)+ \logpow{L^{4}} \Big(
   \mfrac{54}{5}\zeta_2^2
  - \mfrac{1078}{9}\zeta_3
  - \mfrac{661}{18}\zeta_2
  + \mfrac{964631}{1944}
 \Big)+ \logpow{L^{3}} \Big(
  - \mint{136}\zeta_5
  - \mfrac{244}{3}\zeta_3 \zeta_2
  + \mfrac{382}{15}\zeta_2^2
  + \mfrac{130616}{81}\zeta_3
  - \mfrac{51205}{162}\zeta_2
   \notag \\ &
  - \mfrac{13874701}{5832}
 \Big)+ \logpow{L^{2}} \Big(
   \mfrac{4178}{9}\zeta_3^2
  + \mfrac{39076}{945}\zeta_2^3
  - \mfrac{1834}{9}\zeta_5
  - \mfrac{1546}{9}\zeta_3 \zeta_2
  - \mfrac{11875}{54}\zeta_2^2
  - \mfrac{2038430}{243}\zeta_3
  + \mfrac{2509531}{729}\zeta_2
  + \mfrac{721539817}{104976}
 \Big)
  \notag \\ &
 + \logpow{L} \Big(
   \mint{8610}\zeta_7
  - \mint{1968}\zeta_5 \zeta_2
  - \mfrac{3184}{15}\zeta_3 \zeta_2^2
  - \mfrac{42620}{9}\zeta_3^2
  - \mfrac{48266}{315}\zeta_2^3
  + \mfrac{8192}{3}\zeta_5
  + \mfrac{16429}{3}\zeta_3 \zeta_2
  - \mfrac{398749}{810}\zeta_2^2
  + \mfrac{21435407}{972}\zeta_3
  - \mfrac{21743419}{1458}\zeta_2
   \notag \\ &
  - \mfrac{407701829}{34992}
 \Big) + \logpow{}\Big(
   \mint{1046}\zeta_{5,3}
  - \mint{5104}\zeta_5 \zeta_3
  + \mfrac{24208}{9}\zeta_3^2 \zeta_2
  - \mfrac{3829877}{4725}\zeta_2^4
  - \mfrac{248037}{16}\zeta_7
  - \mfrac{6781}{18}\zeta_5 \zeta_2
  + \mfrac{64919}{45}\zeta_3 \zeta_2^2
  + \mfrac{1022996}{81}\zeta_3^2
   \notag \\ &
  - \mfrac{103553}{420}\zeta_2^3
  - \mfrac{1113539}{216}\zeta_5
  - \mfrac{20087587}{972}\zeta_3 \zeta_2
  + \mfrac{95100011}{29160}\zeta_2^2
  - \mfrac{51597389}{2916}\zeta_3
  + \mfrac{2779278167}{104976}\zeta_2
  + \mfrac{9643400117}{839808}
 \Big)
 \bigg]
 \notag \\ &
+ \CS{C_F C_A^3} \bigg[
 \logpow{L^{5}} \Big(
   \mfrac{1331}{270}
 \Big)+ \logpow{L^{4}} \Big(
   \mfrac{121}{9}\zeta_2
  - \mfrac{33803}{324}
 \Big)+ \logpow{L^{3}} \Big(
   \mfrac{484}{15}\zeta_2^2
  - \mfrac{968}{3}\zeta_3
  - \mfrac{694}{27}\zeta_2
  + \mfrac{467107}{486}
 \Big)+ \logpow{L^{2}} \Big(
   \mint{4}\zeta_3^2
  + \mfrac{5008}{105}\zeta_2^3
  - \mfrac{5830}{9}\zeta_5
   \notag \\ &
  - \mint{132}\zeta_3 \zeta_2
  - \mfrac{121}{15}\zeta_2^2
  + \mfrac{32924}{9}\zeta_3
  - \mfrac{8905}{9}\zeta_2
  - \mfrac{3513241}{729}
 \Big)+ \logpow{L} \Big(
  - \mfrac{45511}{12}\zeta_7
  + \mfrac{824}{3}\zeta_5 \zeta_2
  - \mfrac{2066}{15}\zeta_3 \zeta_2^2
  + \mfrac{15059}{9}\zeta_3^2
  + \mfrac{222632}{945}\zeta_2^3
   \notag \\ &
  + \mfrac{101905}{27}\zeta_5
  - \mfrac{6028}{9}\zeta_3 \zeta_2
  - \mfrac{36947}{90}\zeta_2^2
  - \mfrac{854477}{54}\zeta_3
  + \mfrac{1779112}{243}\zeta_2
  + \mfrac{114204817}{8748}
 \Big) + \logpow{}\Big(
  - \mfrac{14161}{30}\zeta_{5,3}
  + \mfrac{21577}{6}\zeta_5 \zeta_3
  - \mfrac{1963}{3}\zeta_3^2 \zeta_2
   \notag \\ &
  + \mfrac{10233079}{15750}\zeta_2^4
  + \mfrac{616417}{144}\zeta_7
  - \mint{397}\zeta_5 \zeta_2
  - \mfrac{19823}{45}\zeta_3 \zeta_2^2
  - \mfrac{845393}{108}\zeta_3^2
  - \mfrac{8189719}{11340}\zeta_2^3
  - \mfrac{8979437}{3240}\zeta_5
  + \mfrac{720313}{108}\zeta_3 \zeta_2
  - \mfrac{283307}{1620}\zeta_2^2
   \notag \\ &
  + \mfrac{32942281}{1458}\zeta_3
  - \mfrac{540427967}{34992}\zeta_2
  - \mfrac{3289233097}{209952}
 \Big)
 \bigg]
+ \CS{\frac{d^{abcd}_F d^{abcd}_A}{N_F}} \bigg[
 \logpow{L^{2}} \Big(
   \mint{96}\zeta_3^2
  + \mfrac{1984}{35}\zeta_2^3
  - \mfrac{880}{3}\zeta_5
  - \mfrac{32}{3}\zeta_3
  + \mint{32}\zeta_2
 \Big)+ \logpow{L} \Big(
   \mint{1742}\zeta_7
    \notag \\ &
  + \mint{512}\zeta_5 \zeta_2
  - \mfrac{368}{5}\zeta_3 \zeta_2^2
  - \mfrac{1672}{3}\zeta_3^2
  + \mfrac{13904}{315}\zeta_2^3
  - \mfrac{920}{9}\zeta_5
  - \mint{896}\zeta_3 \zeta_2
  + \mfrac{112}{15}\zeta_2^2
  - \mfrac{3904}{9}\zeta_3
  - \mfrac{1088}{3}\zeta_2
  + \mint{96}
 \Big) + \logpow{}\Big(
   \mint{260}\zeta_{5,3}
  - \mint{5092}\zeta_5 \zeta_3
   \notag \\ &
  - \mint{16}\zeta_3^2 \zeta_2
  - \mfrac{496766}{525}\zeta_2^4
  + \mint{3518}\zeta_7
  - \mfrac{4744}{3}\zeta_5 \zeta_2
  + \mfrac{6584}{15}\zeta_3 \zeta_2^2
  + \mfrac{39986}{9}\zeta_3^2
  + \mfrac{526496}{945}\zeta_2^3
  - \mfrac{180566}{27}\zeta_5
  + \mfrac{3020}{3}\zeta_3 \zeta_2
  + \mfrac{1220}{9}\zeta_2^2
   \notag \\ &
  + \mfrac{169532}{27}\zeta_3
  + \mfrac{10570}{9}\zeta_2
  - \mfrac{1580}{3}
 \Big)
 \bigg]
+ \CS{n_f C_F^3} \bigg[
 \logpow{L^{7}} \Big(
  - \mfrac{1}{9}
 \Big)+ \logpow{L^{6}} \Big(
   \mfrac{31}{18}
 \Big)+ \logpow{L^{5}} \Big(
  - \mfrac{4}{9}\zeta_2
  - \mfrac{665}{54}
 \Big)+ \logpow{L^{4}} \Big(
  - \mfrac{119}{9}\zeta_3
  + \mfrac{79}{6}\zeta_2
  + \mfrac{29645}{648}
 \Big)
  \notag \\ &
 + \logpow{L^{3}} \Big(
  - \mfrac{169}{45}\zeta_2^2
  + \mfrac{1342}{9}\zeta_3
  - \mfrac{3085}{27}\zeta_2
  - \mfrac{1975}{27}
 \Big)+ \logpow{L^{2}} \Big(
  - \mfrac{1024}{9}\zeta_5
  - \mfrac{466}{9}\zeta_3 \zeta_2
  + \mfrac{20827}{270}\zeta_2^2
  - \mfrac{49697}{81}\zeta_3
  + \mfrac{153121}{324}\zeta_2
  - \mfrac{15028}{243}
 \Big)
  \notag \\ &
 + \logpow{L} \Big(
   \mfrac{664}{3}\zeta_3^2
  - \mfrac{3278}{105}\zeta_2^3
  + \mfrac{5504}{3}\zeta_5
  + \mfrac{4430}{9}\zeta_3 \zeta_2
  - \mfrac{133739}{270}\zeta_2^2
  - \mfrac{281}{9}\zeta_3
  - \mfrac{105101}{108}\zeta_2
  + \mfrac{225611}{648}
 \Big) + \logpow{}\Big(
   \mfrac{2013}{2}\zeta_7
  - \mfrac{1124}{9}\zeta_5 \zeta_2
   \notag \\ &
  - \mfrac{7567}{45}\zeta_3 \zeta_2^2
  - \mfrac{3032}{3}\zeta_3^2
  - \mfrac{20477}{378}\zeta_2^3
  - \mfrac{105215}{18}\zeta_5
  - \mfrac{113617}{81}\zeta_3 \zeta_2
  + \mfrac{3288893}{3240}\zeta_2^2
  + \mfrac{802207}{162}\zeta_3
  + \mfrac{1539611}{1944}\zeta_2
  - \mfrac{1841095}{7776}
 \Big)
 \bigg]
 \notag \\ &
+ \CS{n_f C_F^2 C_A} \bigg[
 \logpow{L^{6}} \Big(
  - \mfrac{88}{81}
 \Big)+ \logpow{L^{5}} \Big(
  - \mfrac{4}{3}\zeta_2
  + \mfrac{538}{27}
 \Big)+ \logpow{L^{4}} \Big(
   \mfrac{124}{9}\zeta_3
  - \mfrac{8}{9}\zeta_2
  - \mfrac{38720}{243}
 \Big)+ \logpow{L^{3}} \Big(
  - \mfrac{20}{3}\zeta_2^2
  - \mfrac{20638}{81}\zeta_3
  + \mfrac{12566}{81}\zeta_2
   \notag \\ &
  + \mfrac{500083}{729}
 \Big)+ \logpow{L^{2}} \Big(
   \mfrac{244}{3}\zeta_5
  + \mfrac{664}{9}\zeta_3 \zeta_2
  + \mfrac{824}{135}\zeta_2^2
  + \mfrac{124628}{81}\zeta_3
  - \mfrac{1743011}{1458}\zeta_2
  - \mfrac{39939557}{26244}
 \Big)+ \logpow{L} \Big(
  - \mfrac{3376}{9}\zeta_3^2
  - \mfrac{1564}{63}\zeta_2^3
  - \mfrac{2560}{3}\zeta_5
   \notag \\ &
  - \mfrac{17930}{27}\zeta_3 \zeta_2
  + \mfrac{223466}{405}\zeta_2^2
  - \mfrac{2214461}{486}\zeta_3
  + \mfrac{5773673}{1458}\zeta_2
  + \mfrac{20234099}{17496}
 \Big) + \logpow{}\Big(
  - \mfrac{1219}{4}\zeta_7
  + \mint{114}\zeta_5 \zeta_2
  + \mfrac{15934}{45}\zeta_3 \zeta_2^2
  + \mfrac{10904}{81}\zeta_3^2
   \notag \\ &
  - \mfrac{808}{105}\zeta_2^3
  + \mfrac{44981}{18}\zeta_5
  + \mfrac{189565}{81}\zeta_3 \zeta_2
  - \mfrac{6376939}{3645}\zeta_2^2
  + \mfrac{25114571}{5832}\zeta_3
  - \mfrac{547858717}{104976}\zeta_2
  + \mfrac{273777229}{419904}
 \Big)
 \bigg]
+ \CS{n_f C_F C_A^2} \bigg[
 \logpow{L^{5}} \Big(
  - \mfrac{121}{45}
 \Big)
  \notag \\ &
 + \logpow{L^{4}} \Big(
  - \mfrac{44}{9}\zeta_2
  + \mfrac{1460}{27}
 \Big)+ \logpow{L^{3}} \Big(
  - \mfrac{88}{15}\zeta_2^2
  + \mint{88}\zeta_3
  - \mfrac{356}{27}\zeta_2
  - \mfrac{25553}{54}
 \Big)+ \logpow{L^{2}} \Big(
   \mfrac{700}{9}\zeta_5
  - \mint{8}\zeta_3 \zeta_2
  - \mfrac{208}{5}\zeta_2^2
  - \mint{962}\zeta_3
  + \mfrac{15914}{27}\zeta_2
   \notag \\ &
  + \mfrac{545491}{243}
 \Big)+ \logpow{L} \Big(
   \mfrac{1186}{9}\zeta_3^2
  - \mfrac{25244}{945}\zeta_2^3
  + \mfrac{836}{27}\zeta_5
  + \mfrac{572}{9}\zeta_3 \zeta_2
  + \mfrac{469}{15}\zeta_2^2
  + \mfrac{351677}{81}\zeta_3
  - \mfrac{810803}{243}\zeta_2
  - \mfrac{66887935}{11664}
 \Big) + \logpow{}\Big(
   \mfrac{19141}{72}\zeta_7
    \notag \\ &
  - \mfrac{127}{3}\zeta_5 \zeta_2
  - \mfrac{6904}{45}\zeta_3 \zeta_2^2
  + \mfrac{4958}{9}\zeta_3^2
  + \mfrac{345871}{11340}\zeta_2^3
  - \mfrac{3862513}{3240}\zeta_5
  - \mfrac{92201}{108}\zeta_3 \zeta_2
  + \mfrac{316999}{540}\zeta_2^2
  - \mfrac{40209899}{5832}\zeta_3
  + \mfrac{213890551}{34992}\zeta_2
   \notag \\ &
  + \mfrac{5309402065}{839808}
 \Big)
 \bigg]
+ \CS{n_f \frac{d^{abcd}_F d^{abcd}_F}{N_F}} \bigg[
 \logpow{L^{2}} \Big(
   \mfrac{320}{3}\zeta_5
  + \mfrac{64}{3}\zeta_3
  - \mint{64}\zeta_2
 \Big)+ \logpow{L} \Big(
   \mfrac{608}{3}\zeta_3^2
  + \mfrac{4736}{315}\zeta_2^3
  - \mfrac{10880}{9}\zeta_5
  + \mint{64}\zeta_3 \zeta_2
  - \mfrac{160}{3}\zeta_2^2
   \notag \\ &
  - \mfrac{2656}{9}\zeta_3
  + \mfrac{2272}{3}\zeta_2
  - \mint{192}
 \Big) + \logpow{}\Big(
  - \mint{1240}\zeta_7
  + \mfrac{992}{3}\zeta_5 \zeta_2
  - \mfrac{3952}{15}\zeta_3 \zeta_2^2
  + \mfrac{680}{9}\zeta_3^2
  + \mfrac{41620}{189}\zeta_2^3
  + \mfrac{95098}{27}\zeta_5
  + \mfrac{92}{3}\zeta_3 \zeta_2
  + \mfrac{7552}{45}\zeta_2^2
   \notag \\ &
  - \mfrac{13414}{27}\zeta_3
  - \mfrac{21566}{9}\zeta_2
  + \mfrac{3190}{3}
 \Big)
 \bigg]
+ \CS{n_{q\gamma} C_F \frac{d^{abc}_F d^{abc}_F}{N_F}} \bigg[
 \logpow{L^{2}} \Big(
   \mfrac{640}{3}\zeta_5
  + \mfrac{16}{5}\zeta_2^2
  - \mfrac{112}{3}\zeta_3
  - \mint{80}\zeta_2
  - \mint{32}
 \Big)+ \logpow{L} \Big(
  - \mint{640}\zeta_5
  - \mfrac{48}{5}\zeta_2^2
   \notag \\ &
  + \mint{112}\zeta_3
  + \mint{240}\zeta_2
  + \mint{96}
 \Big) + \logpow{}\Big(
   \mfrac{11536}{3}\zeta_7
  + \mfrac{1280}{3}\zeta_5 \zeta_2
  + \mfrac{1408}{5}\zeta_3 \zeta_2^2
  - \mint{672}\zeta_3^2
  - \mfrac{25808}{105}\zeta_2^3
  + \mfrac{10160}{3}\zeta_5
  - \mfrac{2672}{3}\zeta_3 \zeta_2
  - \mfrac{1392}{5}\zeta_2^2
   \notag \\ &
  - \mfrac{2752}{3}\zeta_3
  - \mint{1376}\zeta_2
  - \mfrac{7040}{9}
 \Big)
 \bigg]
+ \CS{n_{q\gamma} C_A \frac{d^{abc}_F d^{abc}_F}{N_F}} \bigg[
 \logpow{L} \Big(
   \mfrac{7040}{3}\zeta_5
  + \mfrac{176}{5}\zeta_2^2
  - \mfrac{1232}{3}\zeta_3
  - \mint{880}\zeta_2
  - \mint{352}
 \Big) + \logpow{}\Big(
  - \mfrac{13972}{3}\zeta_7
   \notag \\ &
  - \mint{1840}\zeta_5 \zeta_2
  - \mfrac{784}{5}\zeta_3 \zeta_2^2
  - \mfrac{8752}{3}\zeta_3^2
  - \mfrac{523448}{945}\zeta_2^3
  - \mfrac{11740}{9}\zeta_5
  + \mfrac{7192}{3}\zeta_3 \zeta_2
  - \mfrac{43948}{45}\zeta_2^2
  + \mfrac{12568}{3}\zeta_3
  + \mfrac{39344}{9}\zeta_2
  + \mfrac{20384}{9}
 \Big)
 \bigg]
 \notag \\ &
+ \CS{n_f^2 C_F^2} \bigg[
 \logpow{L^{6}} \Big(
   \mfrac{8}{81}
 \Big)+ \logpow{L^{5}} \Big(
  - \mfrac{44}{27}
 \Big)+ \logpow{L^{4}} \Big(
   \mfrac{10}{9}\zeta_2
  + \mfrac{5741}{486}
 \Big)+ \logpow{L^{3}} \Big(
   \mfrac{380}{81}\zeta_3
  - \mfrac{434}{27}\zeta_2
  - \mfrac{62959}{1458}
 \Big)+ \logpow{L^{2}} \Big(
   \mfrac{692}{135}\zeta_2^2
  - \mfrac{11350}{243}\zeta_3
  + \mfrac{7415}{81}\zeta_2
   \notag \\ &
  + \mfrac{1473913}{26244}
 \Big)+ \logpow{L} \Big(
  - \mfrac{104}{3}\zeta_5
  - \mfrac{568}{27}\zeta_3 \zeta_2
  - \mfrac{184}{3}\zeta_2^2
  + \mfrac{71018}{243}\zeta_3
  - \mfrac{170090}{729}\zeta_2
  + \mfrac{235108}{2187}
 \Big) + \logpow{}\Big(
   \mfrac{4556}{81}\zeta_3^2
  + \mfrac{3520}{189}\zeta_2^3
  + \mfrac{3796}{27}\zeta_5
   \notag \\ &
  + \mfrac{18802}{243}\zeta_3 \zeta_2
  + \mfrac{107507}{810}\zeta_2^2
  - \mfrac{514580}{729}\zeta_3
  + \mfrac{5818805}{26244}\zeta_2
  - \mfrac{73476853}{209952}
 \Big)
 \bigg]
+ \CS{n_f^2 C_F C_A} \bigg[
 \logpow{L^{5}} \Big(
   \mfrac{22}{45}
 \Big)+ \logpow{L^{4}} \Big(
   \mfrac{4}{9}\zeta_2
  - \mfrac{242}{27}
 \Big)+ \logpow{L^{3}} \Big(
  - \mfrac{16}{3}\zeta_3
   \notag \\ &
  + \mfrac{184}{27}\zeta_2
  + \mfrac{11651}{162}
 \Big)+ \logpow{L^{2}} \Big(
   \mfrac{20}{3}\zeta_2^2
  + \mfrac{508}{9}\zeta_3
  - \mfrac{2860}{27}\zeta_2
  - \mfrac{154343}{486}
 \Big)+ \logpow{L} \Big(
  - \mfrac{616}{9}\zeta_5
  + \mfrac{152}{9}\zeta_3 \zeta_2
  + \mfrac{308}{15}\zeta_2^2
  - \mfrac{21604}{81}\zeta_3
  + \mfrac{114943}{243}\zeta_2
   \notag \\ &
  + \mfrac{8996449}{11664}
 \Big) + \logpow{}\Big(
  - \mfrac{622}{27}\zeta_3^2
  + \mfrac{1654}{135}\zeta_2^3
  + \mfrac{22874}{135}\zeta_5
  - \mfrac{956}{27}\zeta_3 \zeta_2
  - \mfrac{18431}{135}\zeta_2^2
  + \mfrac{719659}{1458}\zeta_3
  - \mfrac{26318309}{34992}\zeta_2
  - \mfrac{689230799}{839808}
 \Big)
 \bigg]
 \notag \\ &
+ \CS{n_{q\gamma} n_f \frac{d^{abc}_F d^{abc}_F}{N_F}} \bigg[
 \logpow{L} \Big(
  - \mfrac{1280}{3}\zeta_5
  - \mfrac{32}{5}\zeta_2^2
  + \mfrac{224}{3}\zeta_3
  + \mint{160}\zeta_2
  + \mint{64}
 \Big) + \logpow{}\Big(
   \mfrac{1408}{3}\zeta_3^2
  + \mfrac{11264}{135}\zeta_2^3
  + \mfrac{3520}{9}\zeta_5
  - \mfrac{448}{3}\zeta_3 \zeta_2
  + \mfrac{608}{9}\zeta_2^2
   \notag \\ &
  - \mint{224}\zeta_3
  - \mfrac{4448}{9}\zeta_2
  - \mfrac{3136}{9}
 \Big)
 \bigg]
+ \CS{n_f^3 C_F} \bigg[
 \logpow{L^{5}} \Big(
  - \mfrac{4}{135}
 \Big)+ \logpow{L^{4}} \Big(
   \mfrac{38}{81}
 \Big)+ \logpow{L^{3}} \Big(
  - \mfrac{16}{27}\zeta_2
  - \mfrac{812}{243}
 \Big)+ \logpow{L^{2}} \Big(
   \mfrac{152}{27}\zeta_2
  + \mfrac{9910}{729}
 \Big)
  \notag \\ &
 + \logpow{L} \Big(
  - \mfrac{104}{45}\zeta_2^2
  - \mfrac{40}{81}\zeta_3
  - \mfrac{1624}{81}\zeta_2
  - \mfrac{69874}{2187}
 \Big) + \logpow{}\Big(
  - \mfrac{106}{135}\zeta_5
  + \mfrac{4}{9}\zeta_3 \zeta_2
  + \mfrac{3044}{405}\zeta_2^2
  + \mfrac{104}{243}\zeta_3
  + \mfrac{19766}{729}\zeta_2
  + \mfrac{1865531}{52488}
 \Big)
 \bigg]

%% file: eq/cg.tex
C^g_4 &= \CS{C_A^4} \bigg[
 \logpow{L^{8}} \Big(
   \mfrac{1}{24}
 \Big)+ \logpow{L^{7}} \Big(
   \mfrac{11}{18}
 \Big)+ \logpow{L^{6}} \Big(
   \mfrac{5}{6}\zeta_2
  - \mfrac{119}{162}
 \Big)+ \logpow{L^{5}} \Big(
  - \zeta_3
  + \mfrac{22}{9}\zeta_2
  - \mfrac{6659}{270}
 \Big)+ \logpow{L^{4}} \Big(
   \mfrac{181}{20}\zeta_2^2
  - \mfrac{187}{18}\zeta_3
  - \mfrac{6079}{108}\zeta_2
  + \mfrac{30547}{972}
 \Big)
 \notag \\ &
 + \logpow{L^{3}} \Big(
  - \mint{16}\zeta_5
  - \mfrac{46}{3}\zeta_3 \zeta_2
  - \mfrac{1969}{90}\zeta_2^2
  - \mfrac{6511}{81}\zeta_3
  + \mfrac{6307}{162}\zeta_2
  + \mfrac{391033}{729}
 \Big)+ \logpow{L^{2}} \Big(
   \mfrac{158}{9}\zeta_3^2
  + \mfrac{99791}{1890}\zeta_2^3
  - \mfrac{704}{3}\zeta_5
  + \mfrac{55}{9}\zeta_3 \zeta_2
  - \mfrac{41173}{135}\zeta_2^2
   \notag \\ &
  + \mfrac{37027}{243}\zeta_3
  + \mfrac{761857}{729}\zeta_2
  - \mfrac{1591244}{6561}
 \Big)+ \logpow{L} \Big(
  - \mfrac{2671}{12}\zeta_7
  - \mfrac{400}{3}\zeta_5 \zeta_2
  - \mfrac{307}{5}\zeta_3 \zeta_2^2
  + \mint{143}\zeta_3^2
  - \mfrac{164747}{945}\zeta_2^3
  - \mfrac{63710}{27}\zeta_5
  + \mfrac{26983}{27}\zeta_3 \zeta_2
  \notag \\ &
  + \mfrac{355048}{405}\zeta_2^2
  + \mfrac{1763579}{243}\zeta_3
  - \mfrac{1989215}{729}\zeta_2
  - \mfrac{39280459}{2916}
 \Big)
 + \logpow{}\Big(
  - \mfrac{181}{30}\zeta_{5,3}
  + \mfrac{2377}{6}\zeta_5 \zeta_3
  + \mfrac{271}{9}\zeta_3^2 \zeta_2
  + \mfrac{4583689}{27000}\zeta_2^4
  - \mfrac{224939}{72}\zeta_7
  \notag \\ &
  + \mfrac{5423}{6}\zeta_5 \zeta_2
  + \mfrac{18931}{90}\zeta_3 \zeta_2^2
  + \mfrac{418801}{162}\zeta_3^2
  + \mfrac{353093}{1620}\zeta_2^3
  + \mfrac{1203647}{135}\zeta_5
  - \mfrac{1806605}{486}\zeta_3 \zeta_2
  - \mfrac{778313}{5832}\zeta_2^2
  - \mfrac{47586469}{1944}\zeta_3
  + \mfrac{32379341}{104976}\zeta_2
  \notag \\ &
  + \mfrac{5165679667}{139968}
 \Big)
 \bigg]
+ \CS{\frac{d^{abcd}_A d^{abcd}_A}{N_A}} \bigg[
 \logpow{L^{2}} \Big(
   \mint{96}\zeta_3^2
  + \mfrac{1984}{35}\zeta_2^3
  - \mfrac{880}{3}\zeta_5
  - \mfrac{32}{3}\zeta_3
  + \mint{32}\zeta_2
 \Big)+ \logpow{L} \Big(
   \mint{1742}\zeta_7
  + \mint{512}\zeta_5 \zeta_2
  - \mfrac{368}{5}\zeta_3 \zeta_2^2
  \notag \\ &
  - \mfrac{1672}{3}\zeta_3^2
  + \mfrac{19888}{315}\zeta_2^3
  + \mfrac{1360}{9}\zeta_5
  - \mint{1168}\zeta_3 \zeta_2
  - \mfrac{904}{15}\zeta_2^2
  - \mfrac{14704}{9}\zeta_3
  + \mint{32}\zeta_2
  + \mfrac{128}{3}
 \Big) + \logpow{}\Big(
   \mint{260}\zeta_{5,3}
  - \mint{5092}\zeta_5 \zeta_3
  - \mint{16}\zeta_3^2 \zeta_2
  \notag \\ &
  - \mfrac{496766}{525}\zeta_2^4
  - \mfrac{6776}{3}\zeta_7
  - \mint{5016}\zeta_5 \zeta_2
  + \mfrac{2992}{3}\zeta_3 \zeta_2^2
  + \mfrac{31588}{3}\zeta_3^2
  + \mfrac{1073972}{945}\zeta_2^3
  - \mint{6460}\zeta_5
  + \mfrac{6752}{9}\zeta_3 \zeta_2
  + \mfrac{24616}{45}\zeta_2^2
  + \mfrac{68410}{9}\zeta_3
  \notag \\ &
  - \mfrac{4682}{27}\zeta_2
  - \mfrac{1310}{9}
 \Big)
 \bigg]
+ \CS{n_f C_A^3} \bigg[
 \logpow{L^{7}} \Big(
  - \mfrac{1}{9}
 \Big)+ \logpow{L^{6}} \Big(
  - \mfrac{43}{81}
 \Big)+ \logpow{L^{5}} \Big(
  - \mfrac{4}{9}\zeta_2
  + \mfrac{379}{45}
 \Big)+ \logpow{L^{4}} \Big(
  - \mfrac{91}{9}\zeta_3
  + \mfrac{739}{54}\zeta_2
  + \mfrac{9245}{486}
 \Big)
 \notag \\ &
 + \logpow{L^{3}} \Big(
   \mfrac{323}{45}\zeta_2^2
  - \mfrac{3028}{81}\zeta_3
  + \mfrac{64}{3}\zeta_2
  - \mfrac{426073}{1458}
 \Big)+ \logpow{L^{2}} \Big(
  - \mfrac{808}{9}\zeta_5
  - \mfrac{178}{9}\zeta_3 \zeta_2
  + \mfrac{1202}{15}\zeta_2^2
  + \mfrac{2594}{9}\zeta_3
  - \mfrac{368554}{729}\zeta_2
  - \mfrac{13093133}{26244}
 \Big)
 \notag \\ &
 + \logpow{L} \Big(
  - \mint{46}\zeta_3^2
  + \mfrac{5726}{135}\zeta_2^3
  + \mfrac{7241}{27}\zeta_5
  + \mfrac{1028}{27}\zeta_3 \zeta_2
  - \mfrac{104077}{405}\zeta_2^2
  - \mfrac{445903}{243}\zeta_3
  + \mfrac{645433}{729}\zeta_2
  + \mfrac{347217971}{34992}
 \Big) + \logpow{}\Big(
  - \mfrac{8390}{9}\zeta_7
  \notag \\ &
  + \mfrac{991}{9}\zeta_5 \zeta_2
  - \mfrac{2129}{45}\zeta_3 \zeta_2^2
  - \mfrac{32425}{324}\zeta_3^2
  - \mfrac{702253}{5670}\zeta_2^3
  + \mfrac{566977}{540}\zeta_5
  + \mfrac{67831}{162}\zeta_3 \zeta_2
  - \mfrac{2333729}{29160}\zeta_2^2
  + \mfrac{9686917}{1944}\zeta_3
  + \mfrac{113944685}{104976}\zeta_2
  \notag \\ &
  - \mfrac{20463665839}{839808}
 \Big)
 \bigg]
+ \CS{n_f C_A^2 C_F} \bigg[
 \logpow{L^{5}} \Big(
   \mfrac{5}{3}
 \Big)+ \logpow{L^{4}} \Big(
   \mint{12}\zeta_3
  - \mfrac{23}{12}
 \Big)+ \logpow{L^{3}} \Big(
  - \mfrac{16}{5}\zeta_2^2
  + \mfrac{728}{9}\zeta_3
  - \mfrac{3239}{27}
 \Big)+ \logpow{L^{2}} \Big(
  - \mfrac{968}{9}\zeta_5
  + \mfrac{56}{3}\zeta_3 \zeta_2
  \notag \\ &
  + \mfrac{64}{45}\zeta_2^2
  - \mfrac{9776}{81}\zeta_3
  - \mfrac{1817}{18}\zeta_2
  + \mfrac{33560}{243}
 \Big)+ \logpow{L} \Big(
   \mint{60}\zeta_3^2
  + \mfrac{3824}{315}\zeta_2^3
  - \mfrac{2228}{3}\zeta_5
  - \mfrac{3016}{9}\zeta_3 \zeta_2
  + \mfrac{331}{45}\zeta_2^2
  - \mfrac{50758}{27}\zeta_3
  + \mfrac{9706}{27}\zeta_2
  \notag \\ &
  + \mfrac{7284955}{1944}
 \Big) + \logpow{}\Big(
   \mfrac{16003}{12}\zeta_7
  + \mfrac{230}{9}\zeta_5 \zeta_2
  - \mfrac{44}{15}\zeta_3 \zeta_2^2
  - \mfrac{1787}{3}\zeta_3^2
  + \mfrac{32254}{945}\zeta_2^3
  + \mfrac{143197}{36}\zeta_5
  + \mfrac{78590}{81}\zeta_3 \zeta_2
  - \mfrac{44839}{540}\zeta_2^2
  + \mfrac{8317937}{1944}\zeta_3
  \notag \\ &
  - \mfrac{293267}{3888}\zeta_2
  - \mfrac{573672965}{46656}
 \Big)
 \bigg]
+ \CS{n_f C_A C_F^2} \bigg[
 \logpow{L^{3}} \Big(
   \mfrac{7}{3}
 \Big)+ \logpow{L^{2}} \Big(
   \mint{240}\zeta_5
  - \mint{148}\zeta_3
  - \mfrac{116}{3}
 \Big)+ \logpow{L} \Big(
  - \mint{40}\zeta_3^2
  - \mfrac{160}{7}\zeta_2^3
  + \mfrac{4480}{3}\zeta_5
  + \mfrac{74}{5}\zeta_2^2
  \notag \\ &
  - \mfrac{8084}{9}\zeta_3
  - \mint{4}\zeta_2
  - \mfrac{6827}{216}
 \Big) + \logpow{}\Big(
  - \mfrac{9580}{3}\zeta_7
  - \mint{300}\zeta_5 \zeta_2
  + \mint{12}\zeta_3 \zeta_2^2
  - \mint{368}\zeta_3^2
  - \mfrac{39328}{945}\zeta_2^3
  - \mfrac{92317}{18}\zeta_5
  + \mfrac{193}{3}\zeta_3 \zeta_2
  - \mint{5}\zeta_2^2
  + \mfrac{700879}{108}\zeta_3
  \notag \\ &
  - \mfrac{217}{36}\zeta_2
  + \mfrac{1156175}{1296}
 \Big)
 \bigg]
+ \CS{n_f C_F^3} \bigg[
 \logpow{L} \Big(
  - \mint{69}
 \Big) + \logpow{}\Big(
   \mint{3360}\zeta_7
  - \mint{2940}\zeta_5
  - \mint{156}\zeta_3
  + \mfrac{169}{2}
 \Big)
 \bigg]
+ \CS{n_f \frac{d^{abcd}_A d^{abcd}_F}{N_A}} \bigg[
 \logpow{L^{2}} \Big(
   \mfrac{320}{3}\zeta_5
  + \mfrac{64}{3}\zeta_3
 \notag \\ &
  - \mint{64}\zeta_2
 \Big)+ \logpow{L} \Big(
   \mfrac{608}{3}\zeta_3^2
  - \mfrac{7232}{315}\zeta_2^3
  - \mfrac{15440}{9}\zeta_5
  + \mint{608}\zeta_3 \zeta_2
  + \mfrac{1232}{15}\zeta_2^2
  + \mfrac{21248}{9}\zeta_3
  - \mint{32}\zeta_2
  - \mfrac{608}{3}
 \Big) + \logpow{}\Big(
   \mfrac{2464}{3}\zeta_7
  + \mint{1824}\zeta_5 \zeta_2
  \notag \\ &
  - \mfrac{1088}{3}\zeta_3 \zeta_2^2
  - \mfrac{15700}{3}\zeta_3^2
  - \mfrac{245536}{945}\zeta_2^3
  + \mfrac{108692}{9}\zeta_5
  + \mfrac{1544}{9}\zeta_3 \zeta_2
  - \mfrac{35108}{45}\zeta_2^2
  - \mfrac{89932}{9}\zeta_3
  + \mfrac{9580}{27}\zeta_2
  + \mfrac{6944}{9}
 \Big)
 \bigg]
 \notag \\ &
+ \CS{n_f^2 C_A^2} \bigg[
 \logpow{L^{6}} \Big(
   \mfrac{8}{81}
 \Big)+ \logpow{L^{5}} \Big(
  - \mfrac{34}{135}
 \Big)+ \logpow{L^{4}} \Big(
  - \mfrac{22}{27}\zeta_2
  - \mfrac{1589}{243}
 \Big)+ \logpow{L^{3}} \Big(
   \mfrac{908}{81}\zeta_3
  - \mfrac{602}{81}\zeta_2
  + \mfrac{34133}{1458}
 \Big)+ \logpow{L^{2}} \Big(
  - \mfrac{752}{135}\zeta_2^2
  - \mfrac{3904}{243}\zeta_3
  \notag \\ &
  + \mfrac{626}{9}\zeta_2
  + \mfrac{6162409}{26244}
 \Big)+ \logpow{L} \Big(
   \mfrac{344}{9}\zeta_5
  - \mfrac{320}{9}\zeta_3 \zeta_2
  + \mfrac{1532}{135}\zeta_2^2
  - \mfrac{51139}{243}\zeta_3
  - \mfrac{58751}{729}\zeta_2
  - \mfrac{64308517}{34992}
 \Big) + \logpow{}\Big(
   \mfrac{9452}{81}\zeta_3^2
  + \mfrac{15044}{945}\zeta_2^3
  \notag \\ &
  - \mfrac{38071}{135}\zeta_5
  + \mfrac{3113}{486}\zeta_3 \zeta_2
  + \mfrac{78953}{3240}\zeta_2^2
  + \mfrac{1103621}{1944}\zeta_3
  - \mfrac{25105537}{104976}\zeta_2
  + \mfrac{3255482741}{839808}
 \Big)
 \bigg]
+ \CS{n_f^2 C_A C_F} \bigg[
 \logpow{L^{4}} \Big(
  - \mfrac{7}{3}
 \Big)+ \logpow{L^{3}} \Big(
  - \mfrac{160}{9}\zeta_3
  \notag \\ &
  + \mfrac{547}{27}
 \Big)+ \logpow{L^{2}} \Big(
   \mfrac{16}{45}\zeta_2^2
  - \mfrac{328}{9}\zeta_3
  + \mfrac{116}{9}\zeta_2
  + \mfrac{5701}{54}
 \Big)+ \logpow{L} \Big(
   \mint{152}\zeta_5
  + \mfrac{160}{3}\zeta_3 \zeta_2
  + \mfrac{112}{45}\zeta_2^2
  + \mfrac{20200}{27}\zeta_3
  - \mfrac{64}{3}\zeta_2
  - \mfrac{1469381}{972}
 \Big)
  \notag \\ &
  + \logpow{}\Big(
  - \mint{270}\zeta_3^2
  - \mfrac{10084}{945}\zeta_2^3
  - \mfrac{23572}{27}\zeta_5
  - \mfrac{944}{9}\zeta_3 \zeta_2
  - \mfrac{764}{135}\zeta_2^2
  - \mfrac{724883}{486}\zeta_3
  - \mfrac{4790}{27}\zeta_2
  + \mfrac{48037931}{11664}
 \Big)
 \bigg]
+ \CS{n_f^2 C_F^2} \bigg[
 \logpow{L^{2}} \Big(
   \mint{4}
 \Big)
  \notag \\ &
  + \logpow{L} \Big(
  - \mint{320}\zeta_5
  + \mint{304}\zeta_3
  - \mint{37}
 \Big)+ \logpow{}\Big(
   \mfrac{800}{3}\zeta_3^2
  + \mfrac{13696}{945}\zeta_2^3
  + \mfrac{3920}{3}\zeta_5
  + \mfrac{32}{3}\zeta_3 \zeta_2
  - \mfrac{212}{15}\zeta_2^2
  - \mint{1592}\zeta_3
  + \mfrac{58}{9}\zeta_2
  + \mfrac{32137}{216}
 \Big)
 \bigg]
 \notag \\ &
+ \CS{n_f^2 \frac{d^{abcd}_F d^{abcd}_F}{N_A}} \bigg[
 \logpow{L} \Big(
  - \mint{512}\zeta_3
  + \mfrac{704}{3}
 \Big) + \logpow{}\Big(
   \mint{512}\zeta_3^2
  - \mint{960}\zeta_5
  + \mfrac{384}{5}\zeta_2^2
  + \mint{1520}\zeta_3
  - \mfrac{9008}{9}
 \Big)
 \bigg]
+ \CS{n_f^3 C_A} \bigg[
 \logpow{L^{5}} \Big(
  - \mfrac{4}{135}
 \Big)+ \logpow{L^{4}} \Big(
   \mfrac{20}{81}
 \Big)
  \notag \\ &
  + \logpow{L^{3}} \Big(
   \mfrac{16}{27}\zeta_2
  + \mfrac{232}{243}
 \Big)+ \logpow{L^{2}} \Big(
  - \mfrac{16}{3}\zeta_3
  - \mfrac{80}{27}\zeta_2
  - \mfrac{13985}{729}
 \Big)+ \logpow{L} \Big(
   \mfrac{56}{45}\zeta_2^2
  + \mfrac{2984}{81}\zeta_3
  + \mfrac{64}{81}\zeta_2
  + \mfrac{199244}{2187}
 \Big) + \logpow{}\Big(
  - \mfrac{194}{15}\zeta_5
  + \mfrac{124}{27}\zeta_3 \zeta_2
  \notag \\ &
  - \mfrac{944}{405}\zeta_2^2
  - \mfrac{17818}{243}\zeta_3
  + \mfrac{9430}{729}\zeta_2
  - \mfrac{8399887}{52488}
 \Big)
 \bigg]
+ \CS{n_f^3 C_F} \bigg[
 \logpow{L^{3}} \Big(
   \mfrac{8}{9}
 \Big)+ \logpow{L^{2}} \Big(
   \mfrac{32}{3}\zeta_3
  - \mfrac{52}{3}
 \Big)+ \logpow{L} \Big(
  - \mfrac{32}{45}\zeta_2^2
  - \mfrac{224}{3}\zeta_3
  - \mfrac{40}{9}\zeta_2
  + \mfrac{2936}{27}
 \Big)
  \notag \\ &
 + \logpow{}\Big(
   \mfrac{640}{27}\zeta_5
  - \mfrac{64}{9}\zeta_3 \zeta_2
  + \mfrac{112}{45}\zeta_2^2
  + \mfrac{4060}{27}\zeta_3
  + \mfrac{64}{3}\zeta_2
  - \mfrac{233953}{972}
 \Big)
 \bigg] \, .